
\def\abs#1{\left\vert #1 \right\vert}

\def\ave#1{\left\langle #1 \right\rangle}

\def\part#1#2{{\partial #1\over\partial #2}}

\def\eps{{\epsilon}}

\def\arcsecf {\hbox{$.\!\!^{\prime\prime}$}}
\def\arcminf {\hbox{$.\!\!^{\prime}$}}
\def\vp{\varphi}

\def\Real{{\rm I\mathchoice{\kern-0.70mm}{\kern-0.70mm}{\kern-0.65mm}%
  {\kern-0.50mm}R}}
\def\C{\rm C\kern-.42em\vrule width.03em height.58em depth-.02em
       \kern.4em}
\font \bolditalics = cmmib10
\def\bx#1{\leavevmode\thinspace\hbox{\vrule\vtop{\vbox{\hrule\kern1pt
        \hbox{\vphantom{\tt/}\thinspace{\bf#1}\thinspace}}
      \kern1pt\hrule}\vrule}\thinspace}

\def \vc #1{{\textfont1=\bolditalics \hbox{$\bf#1$}}}
{\catcode`\@=11
\gdef\SchlangeUnter#1#2{\lower2pt\vbox{\baselineskip 0pt \lineskip0pt
  \ialign{$\m@th#1\hfil##\hfil$\crcr#2\crcr\sim\crcr}}}
}
\def\gtrsim{\mathrel{\mathpalette\SchlangeUnter>}}

\def\ueber#1#2{{\setbox0=\hbox{$#1$}%
  \setbox1=\hbox to\wd0{\hss$\scriptscriptstyle #2$\hss}%
  \offinterlineskip
  \vbox{\box1\kern0.4mm\box0}}{}}

\def\bx#1{\leavevmode\thinspace\hbox{\vrule\vtop{\vbox{\hrule\kern1pt
        \hbox{\vphantom{\tt/}\thinspace{\bf#1}\thinspace}}
      \kern1pt\hrule}\vrule}\thinspace}

\def\SFB{{This work was supported by the ``Sonderforschungsbereich
375-95 f\"ur
Astro--Teil\-chen\-phy\-sik" der Deutschen For\-schungs\-ge\-mein\-schaft
and by the Programme National de Cosmologie
of the Centre National de la Recherche Scientifique in France. We
thank S.D.M. White for carefully reading the manuscript.}}
\voffset=0pt

\magnification=\magstep1
\input epsf
\voffset= 0.0 true cm
\vsize=19.8 cm     
\hsize=13.5 cm
\hfuzz=2pt
\tolerance=500
\abovedisplayskip=3 mm plus6pt minus 4pt
\belowdisplayskip=3 mm plus6pt minus 4pt
\abovedisplayshortskip=0mm plus6pt
\belowdisplayshortskip=2 mm plus4pt minus 4pt
\predisplaypenalty=0
\footline={\tenrm\ifodd\pageno\hfil\folio\else\folio\hfil\fi}

\def\la{\mathrel{\hbox{\rlap{\hbox{\lower4pt\hbox{$\sim$}}}\hbox{$<$}}}}
\def\ga{\mathrel{\hbox{\rlap{\hbox{\lower4pt\hbox{$\sim$}}}\hbox{$>$}}}}

\def\utw{\smash{\rlap{\lower5pt\hbox{$\sim$}}}}
\def\udtw{\smash{\rlap{\lower6pt\hbox{$\approx$}}}}

\def\getsto{\mathrel{\hbox{\rlap{$\gets$}\hbox{\raise2pt\hbox{$\to$}}}}}
\def\lid{\mathrel{\hbox{\rlap{\hbox{\lower4pt\hbox{$=$}}}\hbox{$<$}}}}
\def\gid{\mathrel{\hbox{\rlap{\hbox{\lower4pt\hbox{$=$}}}\hbox{$>$}}}}
\def\sol{\mathrel{\hbox{\rlap{\hbox{\raise4pt\hbox{$\sim$}}}\hbox{$<$}}}
}
\def\sog{\mathrel{\hbox{\rlap{\hbox{\raise4pt\hbox{$\sim$}}}\hbox{$>$}}}
}
\def\lse{\mathrel{\hbox{\rlap{\hbox{\raise4pt\hbox{$<$}}}\hbox{$\simeq$}
}}}
\def\gse{\mathrel{\hbox{\rlap{\hbox{\raise4pt\hbox{$>$}}}\hbox{$\simeq$}
}}}
\def\grole{\mathrel{\hbox{\lower2pt\hbox{$<$}}\kern-8pt
\hbox{\raise2pt\hbox{$>$}}}}
\def\leogr{\mathrel{\hbox{\lower2pt\hbox{$>$}}\kern-8pt
\hbox{\raise2pt\hbox{$<$}}}}
\def\loa{\mathrel{\hbox{\rlap{\hbox{\lower4pt\hbox{$\approx$}}}\hbox{$<$
}}}}
\def\goa{\mathrel{\hbox{\rlap{\hbox{\lower4pt\hbox{$\approx$}}}\hbox{$>$
}}}}

%
%

\font\kleinhalbcurs=cmmib10 scaled 833
\font\eightrm=cmr8
\font\sixrm=cmr6
\font\eighti=cmmi8
\font\sixi=cmmi6
\skewchar\eighti='177 \skewchar\sixi='177
\font\eightsy=cmsy8
\font\sixsy=cmsy6
\skewchar\eightsy='60 \skewchar\sixsy='60
\font\eightbf=cmbx8
\font\sixbf=cmbx6
\font\eighttt=cmtt8
\hyphenchar\eighttt=-1
\font\eightsl=cmsl8
\font\eightit=cmti8

\font\bxf=cmbx10
  \mathchardef\Gamma="0100
  \mathchardef\Delta="0101
  \mathchardef\Theta="0102
  \mathchardef\Lambda="0103
  \mathchardef\Xi="0104
  \mathchardef\Pi="0105
  \mathchardef\Sigma="0106
  \mathchardef\Upsilon="0107
  \mathchardef\Phi="0108
  \mathchardef\Psi="0109
  \mathchardef\Omega="010A
\def\rahmen#1{\vskip#1truecm}
\def\begfig#1cm#2\endfig{\par
\setbox1=\vbox{\rahmen{#1}#2}%
\dimen0=\ht1\advance\dimen0by\dp1\advance\dimen0by5\baselineskip
\advance\dimen0by0.4true cm
\ifdim\dimen0>\vsize\pageinsert\box1\vfill\endinsert
\else
\dimen0=\pagetotal\ifdim\dimen0<\pagegoal
\advance\dimen0by\ht1\advance\dimen0by\dp1\advance\dimen0by1.4true cm
\ifdim\dimen0>\vsize
\topinsert\box1\endinsert
\else\vskip1true cm\box1\vskip4true mm\fi
\else\vskip1true cm\box1\vskip4true mm\fi\fi}
\def\figure#1#2{\smallskip\setbox0=\vbox{\noindent\petit{\bf Fig.\ts#1.\
}\ignorespaces #2\smallskip
\count255=0\global\advance\count255by\prevgraf}%
\ifnum\count255>1\box0\else
\centerline{\petit{\bf Fig.\ts#1.\ }\ignorespaces#2}\smallskip\fi}

\def\xfigure#1#2#3#4{\midinsert\noindent
    $$\epsfxsize=#4truecm\epsffile{#3}$$
    \figure{#1}{#2}\endinsert}
\def\xxfigure#1#2#3#4#5#6{\midinsert\noindent
    $$\epsfxsize=#5truecm\epsffile{#4}\epsfxsize=#6truecm\epsffile{#3}$$
    \figure{#1}{#2}\endinsert}

\def\tabcap#1#2{\smallskip\vbox{\noindent\petit{\bf Table\ts#1\unskip.\
}\ignorespaces #2\smallskip}}
\def\begtab#1cm#2\endtab{\par
\ifvoid\topins\midinsert\vbox{#2\rahmen{#1}}\endinsert
\else\topinsert\vbox{#2\kern#1true cm}\endinsert\fi}
\def\rahmen#1{\vskip#1truecm}
\def\begpet{\vskip6pt\bgroup\petit}
\def\endpet{\vskip6pt\egroup}
\def\begref{\par\bgroup\petit
\let\it=\rm\let\bf=\rm\let\sl=\rm\let\INS=N}
\def\petit{\def\rm{\fam0\eightrm}%
\textfont0=\eightrm \scriptfont0=\sixrm \scriptscriptfont0=\fiverm
 \textfont1=\eighti \scriptfont1=\sixi \scriptscriptfont1=\fivei
 \textfont2=\eightsy \scriptfont2=\sixsy \scriptscriptfont2=\fivesy
 \def\it{\fam\itfam\eightit}%
 \textfont\itfam=\eightit
 \def\sl{\fam\slfam\eightsl}%
 \textfont\slfam=\eightsl
 \def\bf{\fam\bffam\eightbf}%
 \textfont\bffam=\eightbf \scriptfont\bffam=\sixbf
 \scriptscriptfont\bffam=\fivebf
 \def\tt{\fam\ttfam\eighttt}%
 \textfont\ttfam=\eighttt
 \normalbaselineskip=9pt
 \setbox\strutbox=\hbox{\vrule height7pt depth2pt width0pt}%
 \normalbaselines\rm
\def\vec##1{\setbox0=\hbox{$##1$}\hbox{\hbox
to0pt{\copy0\hss}\kern0.45pt\box0}}}%
\let\ts=\thinspace
%
\font \tafontt=     cmbx10 scaled\magstep2
\font \tafonts=     cmbx7  scaled\magstep2
\font \tafontss=     cmbx5  scaled\magstep2
\font \tamt= cmmib10 scaled\magstep2
\font \tams= cmmib10 scaled\magstep1
\font \tamss= cmmib10
\font \tast= cmsy10 scaled\magstep2
\font \tass= cmsy7  scaled\magstep2
\font \tasss= cmsy5  scaled\magstep2
\font \tasyt= cmex10 scaled\magstep2
\font \tasys= cmex10 scaled\magstep1
\font \tbfontt=     cmbx10 scaled\magstep1
\font \tbfonts=     cmbx7  scaled\magstep1
\font \tbfontss=     cmbx5  scaled\magstep1
\font \tbst= cmsy10 scaled\magstep1
\font \tbss= cmsy7  scaled\magstep1
\font \tbsss= cmsy5  scaled\magstep1

\newbox\chsta\newbox\chstb\newbox\chstc
\def\centerpar#1{{\advance\hsize by-2\parindent
\rightskip=0pt plus 4em
\leftskip=0pt plus 4em
\parindent=0pt\setbox\chsta=\vbox{#1}%
\global\setbox\chstb=\vbox{\unvbox\chsta
\setbox\chstc=\lastbox
\line{\hfill\unhbox\chstc\unskip\unskip\unpenalty\hfill}}}%
\leftline{\kern\parindent\box\chstb}}
 \def \chap#1{
    \vskip24pt plus 6pt minus 4pt
    \bgroup
 \textfont0=\tafontt \scriptfont0=\tafonts \scriptscriptfont0=\tafontss
 \textfont1=\tamt \scriptfont1=\tams \scriptscriptfont1=\tamss
 \textfont2=\tast \scriptfont2=\tass \scriptscriptfont2=\tasss
 \textfont3=\tasyt \scriptfont3=\tasys \scriptscriptfont3=\tenex
     \baselineskip=18pt
     \lineskip=18pt
     \raggedright
     \pretolerance=10000
     \noindent
     \tafontt
     \ignorespaces#1\vskip7true mm plus6pt minus 4pt
     \egroup\noindent\ignorespaces}%
 \def \sec#1{
     \vskip25true pt plus4pt minus4pt
     \bgroup
 \textfont0=\tbfontt \scriptfont0=\tbfonts \scriptscriptfont0=\tbfontss
 \textfont1=\tams \scriptfont1=\tamss \scriptscriptfont1=\kleinhalbcurs
 \textfont2=\tbst \scriptfont2=\tbss \scriptscriptfont2=\tbsss
 \textfont3=\tasys \scriptfont3=\tenex \scriptscriptfont3=\tenex
     \baselineskip=16pt
     \lineskip=16pt
     \raggedright
     \pretolerance=10000
     \noindent
     \tbfontt
     \ignorespaces #1
     \vskip12true pt plus4pt minus4pt\egroup\noindent\ignorespaces}%
 \def \subs#1{
     \vskip15true pt plus 4pt minus4pt
     \bgroup
     \bxf
     \noindent
     \raggedright
     \pretolerance=10000
     \ignorespaces #1
     \vskip6true pt plus4pt minus4pt\egroup
     \noindent\ignorespaces}%
 \def \subsubs#1{
     \vskip15true pt plus 4pt minus 4pt
     \bgroup
     \bf
     \noindent
     \ignorespaces #1\unskip.\ \egroup
     \ignorespaces}
\def\footnoterule{\kern-3pt\hrule width 2true cm\kern2.6pt}
\newcount\footcount \footcount=0
\def\advftncnt{\advance\footcount by1\global\footcount=\footcount}
\def\fonote#1{\advftncnt$^{\the\footcount}$\begingroup\petit
       \def\textindent##1{\hang\noindent\hbox
       to\parindent{##1\hss}\ignorespaces}%
\vfootnote{$^{\the\footcount}$}{#1}\endgroup}

\newcount\sterne
\outer\def\byebye{\bigskip\typeset
\sterne=1\ifx\speciali\undefined\else
\bigskip Special caracters created by the author
\loop\smallskip\noindent special character No\number\sterne:
\csname special\romannumeral\sterne\endcsname
\advance\sterne by 1\global\sterne=\sterne
\ifnum\sterne<11\repeat\fi
\vfill\supereject\end}
\def\typeset{\centerline{\petit This article was processed by the author
using the \TeX\ Macropackage from Springer-Verlag.}}
 
\chap{Detection of shear due to weak lensing by large-scale structure}

\centerline{P. Schneider$^1$, L. van Waerbeke$^{1,4}$, Y. Mellier$^{2,3}$,}
\centerline{B. Jain$^1$, S. Seitz$^1$ \& B. Fort$^3$}
\bigskip
\noindent
$^1$ Max-Planck-Institut f\"ur Astrophysik, Postfach 1523, D-85740
Garching, Germany, \hfill\break
$^2$ Institut d'Astrophysique de Paris, 98 bis boulevard Arago,
F-75014 Paris, France, \hfill\break
$^3$ DEMIRM, Observatoire de Paris, 61 Avenue de l'Observatoire,
F-75014 Paris, France,\hfill\break
$^4$ Observatoire Midi-Pyr\'en\'ees, 14 Avenue Edouard Belin,
F-31400 Toulouse, France,\hfill\break

\bigskip

\sec{Abstract}
We present evidence for a coherent shear signal in a field containing
the $z=1.2$ radio-source PKS1508-05. Since there were no intervening
mass concentrations known before targeting this field, we interpret
this signal as due to weak lensing 
by large-scale structure. This result is the outcome of a re-analysis
of the observations of Fort et al. (1996) in the fields of three high
redshift QSO/radio-sources. Several tests of the robustness of the
signal were performed: subdivision of the field and tests using
simulated data with randomized ellipticity orientation; using the
pixel-to-pixel autocorrelation function method to measure the shear;
using the correlation function of the galaxy ellipticities with
bootstrap resampling to estimate errors and to find the typical
angular separation that dominates the signal. We find that the field
of PKS1508-05 contains a robust shear signal that is coherent over the
$2'$ by $2'$ field. The signal is detected at a high level of
significance ($\gtrsim 99.6\%$) using the tests described above. The
amplitude of the shear on a scale of $1'$ is about $3\%$, which is
consistent with theoretical expectations from weak lensing by
large-scale structure.

\vfill\eject

\sec{1. Introduction}
Weak gravitational lensing has found its main application up to now in
reconstructing the projected surface mass density in massive clusters
of galaxies (e.g., Kaiser \& Squires 1993; Fahlman et al. 1994; Seitz
et al. 1996; Squires et al. 1996; for recent reviews, see Fort \&
Mellier 1994; Schneider 1996; Narayan \& Bartelmann 1996). Blandford
et al. (1991), Kaiser (1992, 1996), Miralda-Escude (1991),
Villumsen (1996a) and others considered the shear signal caused by the
tidal field of the large-scale matter distribution in the universe.
The two-point correlation function of the observable shear, for example,
can be directly written in terms of the power spectrum of the
cosmological density fluctuations. These investigations, which
are mainly based on the linear theory of the evolution of density
fluctuations, concluded that the rms of the expected shear is of the
order of one percent, depending on the cosmological model, the
normalization of the power spectrum, and the assumed redshift
distribution of the faint galaxies. 

Since the expected signal is so small, 
it is evident that this so-called cosmic shear will be difficult to
observe with ground-based telescopes -- the anisotropy of the PSF and
the instrumental image distortions have to be understood to a percent
level in order to exclude systematic effects. The observational study
by Mould et al. (1994) did not find a shear signal and put an upper
limit on the shear in their field of $\sim 4$\%, though a later
reanalysis of the same data by Villumsen (1996b) yielded a detection
with large formal significance.  Such investigations push the 
instruments and the telecopes to the limits of their capabilities and 
ideally they should be done preferably with the  HST.  Unfortunately, this  
is prohibitively difficult due the small field-of-view of the current
generation of its instruments.

Recently, progess in the theoretical predictions 
has been made by dropping the approximation of
linear density growth. Bernardeau, Van Waerbeke \& Mellier (1996) have
calculated the skewness of the magnification in the weakly non-linear
regime, and Jain \& Seljak (1996) have calculated the rms of the mean
cosmic shear and its two-point correlation function 
using the fully non-linear evolution of the dark matter power spectrum. 
The latter study showed that on scales below $\sim 20$\ts arcmin the cosmic
shear is significantly larger than estimated from linear density
growth, a trend already hinted at in the numerical simulations
presented in Blandford et al. (1991). These new results are
encouraging, and suggest that the detection of cosmic shear 
with ground-based telescopes should be more feasible on small scales. 

In this paper we re-analyze the observations of Fort et al. (1996;
hereafter FMDBK) of image distortions of faint galaxies around
high-redshift QSOs. The goal of their study was 
to test the hypothesis initially proposed by 
Bartelmann \& Schneider (1992) that the
observed association of high-redshift QSOs with foreground galaxies on
arcminute scales
(e.g., Tyson 1986; Fugmann 1990; Bartelmann \& Schneider 1994; Ben\'\i
tez \& Mart\'\i nez-Gonz\'ales 1997, and references therein) is
caused by lensing by large-scale matter inhomogeneities in which the
galaxies are embedded (Bartelmann 1995). Surprisingly, FMDBK 
found a significant shear signal around several of the observed QSOs;
they demonstrated that this shear signal is considerably stronger
than the instrumental image distortion and the anisotropy of the PSF.  
Their results were confirmed at least for the quasar 3C 336 (referenced
as Q1622 in FMDBK) from analyses of  HST images  (Bower \& Smail 1997),
which gives good confidence that the signal detected by FMDBK is real 
and significant.

In this paper, we argue that FMDBK have
detected a cosmic shear signal --- though along biased lines-of-sight,
if the magnification bias hypothesis for the high-redshift QSOs is
correct. In Sect.\ts 2 we briefly summarize the observations and
the results of FMDBK. The shear measurements are then analyzed in
Sect.\ts 3, by comparing them with those obtained from randomizing the
orientation angles of the faint (background) galaxies, thus
determining the significance of the measured shear. In particular it
will be shown, by considering the measured ellipticity of stars in
the fields observed with the SUSI camera at the ESO NTT, that the
anisotropy of the PSF is well below 1\% and thus does not cause a
spurious shear signal. One of the fields, that around the QSO
PKS1508$-$05, exhibits a very strong and nearly uniform shear across the
whole SUSI field ($\sim 2'\times 2'$). In Sect.\ts 4, we present
further evidence for this shear, by calculating the pixel-to-pixel 
auto-correlation function (ACF) of the image, using different cuts for the
brighter images which are excluded in this ACF, thus following the
method introduced by van Waerbeke et al.\ts (1997). This provides 
 good confidence that the shear detected is not an artefact which results from
the specific technique we use to measure the galaxy ellipticity. 
 Finally, a different
measure for the shear is obtained from the two-point correlation
function of galaxy ellipticities; this is constructed in Sect.\ts 5 and
its statistical significance is tested both with bootstrapping methods
as well as with randomizations of position angles. Our results are
discussed in Sect.\ts 6.

\sec{2. Observations and previous results}

The five QSO fields in FMDBK were selected on the basis of their high
radio and optical luminosity. This selection was designed 
to test the hypothesis that
such QSOs are magnification biased by intervening large-scale matter
inhomogeneities. Four of these QSO fields were observed with the SUSI
camera at the ESO New Technology Telescope at La Silla, and one was
observed with FOCAM (3C 336) at the Canada-France-Hawaii-Telescope 
(CFHT) on Mauna Kea. For details of the observing runs, see FMDBK.

The field of PKS1741$-$03 was chosen for studying the instrumental
distortion properties of SUSI; it is crowded by stars and has been
used in FMDBK to demonstrate the remarkable 
stability of the PSF across the SUSI field.

The QSO 3C 336 was observed at CFHT and shows a strong shear signal. A
deep WFPC2 image from the Hubble Space Telescope and extensive
spectroscopy of galaxies in the field (Steidel et al. 1996) has shown
that the distribution of galaxies spans a wide range in redshift; in
fact, the QSO itself appears to reside in a cluster. It is therefore
unclear whether the observed shear, whose presence was confirmed from
the HST image (M.\ Dickinson, private communication; Bower \& Smail
1997) is due to material in the foreground of the QSO, which is
therefore able to magnify it, or due to a mass overdensity surrounding
the QSO. The latter possibility would be no less exciting than the
first and would add another high-redshift cluster to the list of those
which are strong enough for a shear detection (see Luppino \& Kaiser
1997; G.\ Luppino, private communication), implying either that a
substantial fraction of the population of faint galaxies are at
redshifts well beyond one, or that clusters at redshift $z\sim 0.8 -
1$ are very massive, contradicting predictions made in some cosmogonic
models, or both.

Since the stability of the PSF across the SUSI field allows us to
estimate the significance of a measured shear without correcting for
an anisotropic PSF component (e.g., Kaiser, Squires \& Broadhurst
1995), we shall concentrate in this paper on the three QSO fields
PKS0135$-$247, PKS1508$-$05 and 3C446. The seeing in these fields
ranges from $0\arcsecf66$ to $0\arcsecf76$, and the total exposure
time from 13500\ts s to 19700\ts s in the V band. For each QSO field
we obtained two catalogues of objects. One was constructed using the
same method as described in Bonnet \& Mellier (1995), where the object
detection is performed using a standard detection algorithm similar to
FOCAS, and the ellipticity of the images is measured by a weighted
second-order moment scheme. We refer to the corresponding
ellipticities as `Bonnet/Mellier (or BM) ellipticities'. A second
catalog was constructed using the SExtractor software package (Bertin
1996, Bertin \& Arnouts 1996). These ellipticities are determined from
second moments calculated within a fixed limiting isophote of the
objects and will be referred to as `S-ellipticities'. Only objects
with flag $\le 4$ and size $\ge 8$ pixels are included in the
S-catalog. The final object catalogues consist of $(N_{\rm
BM}=148/N_{\rm S}=148)$ objects for 0135, (145/144) for 1508, and
(211/143) for 3C446.

From the S-catalog, we obtained a list of stellar candidates. Each of
these was carefully checked by eye and removed from the list of
stellar objects if they were saturated, not isolated, too faint, or
appeared non-stellar by eye-inspection. The remaining list of `good'
stars contained 4, 5 and 7 objects for 0135, 1508 and 3C446,
respectively. Their ellipticities are plotted in Figs.\ts 1 through
3. We have not attempted here to measure the ellipticity of stars with
the BM-method because this method is designed to avoid as much as
possible the influence of the PSF.

The BM-ellipticities have smaller values than the ellipticities
measured by SExtractor (or any standard algorithms), as was noted in
Bonnet \& Mellier (1995); this is due to the particular weighting
function employed by BM, specifically designed to avoid the core of
the PSF.
There, a simulated image was analyzed, with
a known value of the applied shear and with the same samplings and
seeings as the CFHT and NTT images. The ratio of the mean ellipticity
in this image (adopted to a deep CFHT image of the cluster CL 0024+16)
to the true value of the shear was found to be $\sim 6$. On the other
hand, Wilson, Cole \& Frenk (1996) have performed similar simulations
using FOCAS ellipticities, and they found a typical ratio of $\sim
1.5$ between mean ellipticity and true shear for seeing conditions
which apply to the images investigated here.

In order to relate BM-ellipticities to S-ellipticities, we have
calculated for each of the three QSO fields a complex relative
correction factor $C$, defined such as to minimize
$$
\sum_{i=1}^N \abs{\eps_i^{\rm S}-C \eps_i^{\rm BM}}^2\; ,
\eqno (1)
$$
where the sum extends over all objects which are contained in both
catalogues (with a maximum positional difference of 2 pixels or
$0\arcsecf 26$) and $\eps^{\rm S}$ and $\eps^{\rm BM}$ are the complex
ellipticities [defined such that for an image with elliptical
isophotes of axis ratio $r\le 1$, $\abs{\eps}=(1-r)/(1+r)$]. Ideally,
the resulting value of $C$ should have a very small imaginary part. We
find for the three fields: $C=3.17-0.36{\rm i}$ for 0135,
$C=3.36-0.29{\rm i}$ for 1508, and $C=2.93+0.17{\rm i}$ for 3C446. We
note, however, that the precise values of $C$ depend fairly
sensitively on cuts applied to the S-catalog. From simulations (van
Waerbeke 1997) it became evident that the SExtractor ellipticities
depend on the parameters selected for the detection prior to run
SExtractor. Since the choice of these parameters are to some degree
arbitrary, they are less useful
to us for measuring shear on very faint galaxy images than the BM
algorithm.  This simply reflects that, contrary to BM, SExtractor was
not constructed for measuring such small distortions and should
therefore not be used on ground-based images such as those at hand
(Bertin, private communication). In the following, we will exclusively
use the BM-ellipticities for the very faint galaxies; however, to plot
the mean ellipticities from the BM method on the same graph as the
ellipticities of stars as measured by SExtractor, we will multiply in
these plots the value of $\ave{\eps^{\rm BM}}$ by the real part of
$C$. Note that this restriction in the use of SExtractor does not
apply to section 4 since for the ACF the very faint galaxies present
in the fields are not used.

\sec{3. The significance of the shear}
The main purpose of this paper is to investigate the statistical
significance of the detection of shear in the field of the three QSOs.
As discussed below, it is more difficult to determine and interpret
the precise value of the shear. 

\xfigure{1}{For the field of the QSO PKS0135$-$247, the mean
ellipticity evaluated over
the whole field (solid triangle), over the four subfields with sidelength
1/2 that of the whole field (solid squares), and over the nine
subfields with sidelength 1/3 of the whole field (open hexagons) are
plotted. The mean ellipticity is obtained from the BM-ellipticity,
multiplied by the real part of the correction factor $C$ in order to
allow a comparison with the ellipticity of the stars (crosses)
measured by SExtractor. Circles of constant $\abs{\eps}$ are plotted
to guide the eye}{fig1.tps}{10}

\xfigure{2}{Same as Fig.\ts1, for the field of QSO
PKS1508$-$05}{fig2.tps}{10} 

\xfigure{3}{Same as Fig.\ts1, for the field of QSO
3C446}{fig3.tps}{10} 

In order to test whether the mean ellipticity of images in a
(sub)field is statistically significant, we have performed simulations
by randomizing the position angles of all objects and measuring the
mean ellipticity of these randomized images. For each of the three QSO
fields, $10^4$ such randomized realizations were conducted. For a
given (sub)field, the fraction $f$ of randomized realizations
was then determined in which the absolute value of the 
mean image ellipticity thus obtained 
is larger than the mean ellipticity measured
from the data. Therefore, $f$ is the probability that the measured mean
shear is obtained from a random distribution of image ellipticities with
the same amplitudes as the observed
images. Small values of $f$ indicate that the observed mean shear
is very unlikely to be caused by a random alignment of galaxy
images. We shall call $f$ the error level henceforth.

For each QSO field, we have calculated $f$ for the whole field
(denoted `T'), the four subfields obtained by deviding the total field
into four equally sized squares (denoted `A' -- lower left, `B' --
upper left, `C' -- lower right, `D' -- upper right, in the frames
shown in FMDBK), and the nine subfields obtained by dividing the field
into nine equally sized squares, denoted `1' -- `9', with the
following configuration
$$
\pmatrix{3& 6 &9 \cr
2 & 5 & 8 \cr
1 & 4 & 7 \cr }.
$$
The
corresponding values of $f$ are listed in Tables 1 -- 3 for each of
these (sub)fields, together with the corresponding number $N$ of
objects in these fields and the mean ellipticity, which is the
BM-ellipticity multiplied by the real part of the correction factor
$C$. Note that the error level $f$ is independent of any scaling of $\eps$. 

 From the values in the tables it is clear that the field around QSO
1508 shows the clearest sign of significant shear, whereas for the two
other QSO fields, the evidence for a statistically significant shear
is considerably weaker. The shear seems to
be quite homogeneous across the field of QSO 1508, and the error level
is estimated to be only $\sim 0.2$\%. In addition, for three of the
four subfields of sidelength 1/2 that of the whole field, the error
level is well below 10\%. In the field of QSO 0135, only two subfields
achieve error levels below 3\%, and only one subfield in 3C446 is
statistically significant, though with an error level of $\sim
0.5$\%. Note that the error level does not strictly correlate with the
measured value of $\abs{\ave{\eps}}$; the number $N$ of galaxies per
subfield and their ellipticity distribution are decisive factors in
the determination of the statistical significance.

In order to see whether the measured ellipticities are affected by an
anisotropic contribution to the PSF, we have plotted the mean values
of the ellipticities as given in the tables, together with the
measured ellipticity of the stars in each field in Figs.\ts 1 -- 3. As
can be clearly seen in these figures, the ellipticities of most stars
are very similar, and those stars for which $\eps$ is significantly
different from the rest are probably not isolated stars but are
contaminated by a nearby or underlying object too faint to be
discovered by eye. In addition it should be noted that the mean
ellipticity of stars lies in a direction different from the mean
ellipticity of galaxies in the fields with smallest error levels --
see in particular Fig.\ts 2. We can thus safely exclude the
possibility that the high statistical significance obtained in some
(sub)fields is due to an anisotropic PSF.

\xfigure{4}{For the subfield 6 in the field of 3C446, the
BM-ellipticities multiplied by the real part of the correction factor
$C$ are plotted, together with the SExtractor ellipticities of the
seven stars in this field. The anisotropic distribution leading to a
significant mean ellipticity in this field is easily seen}{fig4.tps}{10}

Whereas the evidence for significant shear in the whole field of 3C446
and PKS 0135 is weak, there are nevertheless some subfields around
these QSOs where significant shear is detected. As an example, we
consider subfield `6' in 3C446, where the error level for the shear
detection is a mere 0.5\%. In Fig.\ts 4, we have plotted the
BM-ellipticities, multiplied by the real part of the correction factor
$C$, of the 26 galaxies in this subfield, together with the 7 stars in
this field. The statistically significant shear can be seen by eye,
the ellipticities are concentrated towards the upper left corner of
the diagram, and it is easily seen that the mean shear in this
subfield, as well as the statistical significance, is not 
dominated by one or two galaxy images. We have checked this
quantitatively, by removing the two galaxies with largest
ellipticities; the resulting value of the error level does not change
significantly.

\medskip
\tabcap{1}{For the QSO 0135 and each (sub)field as described in the
text, the number $N$ of 
objects, the error level $f$, and the measured mean
BM-ellipticity, multiplied by the real part of the correction factor
(see text) are given}
\smallskip
\settabs\+ (Sub)field \quad &\qquad N\qquad &\quad $f$(in \%)\quad
&\quad $\eps_1$ (in \%)\quad & \quad
$\eps_2$ (in \%) \quad &\cr
\+ \hfill (Sub)field  & \hfill N &\hfill $f$(in \%) &\hfill $\eps_1$ (in \%) &
$\hfill \eps_2$ (in \%) &\cr
\smallskip
\+ \hfill T&\hfill  148&\hfill    46.47&\hfill     0.34&\hfill
0.90&\cr
\smallskip
\+ \hfill A&\hfill   38&\hfill    83.26&\hfill  $-$0.58&\hfill     0.69&\cr
\+ \hfill B&\hfill   40&\hfill    39.44&\hfill     1.49&\hfill  $-$0.08&\cr
\+ \hfill C&\hfill   43&\hfill    69.95&\hfill  $-$0.22&\hfill  $-$1.29&\cr
\+ \hfill D&\hfill   27&\hfill     2.38&\hfill     0.82&\hfill
6.16&\cr
\smallskip
\+ \hfill 1&\hfill   14&\hfill    25.30&\hfill  $-$3.18&\hfill     3.99&\cr
\+ \hfill 2&\hfill   21&\hfill    45.31&\hfill     1.59&\hfill     1.09&\cr
\+ \hfill 3&\hfill   16&\hfill    60.32&\hfill  $-$1.00&\hfill  $-$1.65&\cr
\+ \hfill 4&\hfill   33&\hfill    93.39&\hfill  $-$0.58&\hfill  $-$0.17&\cr
\+ \hfill 5&\hfill   12&\hfill    49.99&\hfill     0.33&\hfill     2.98&\cr
\+ \hfill 6&\hfill   18&\hfill     2.28&\hfill     2.72&\hfill     5.82&\cr
\+ \hfill 7&\hfill   14&\hfill    26.39&\hfill     3.48&\hfill  $-$2.96&\cr
\+ \hfill 8&\hfill   12&\hfill    36.37&\hfill  $-$0.41&\hfill  $-$4.11&\cr
\+ \hfill 9&\hfill    8&\hfill    58.69&\hfill     0.05&\hfill     4.65&\cr

\vfill\eject

\medskip
\tabcap{2}{Same as Table 1, for the QSO 1508}
\smallskip
\settabs\+ (Sub)field \quad &\qquad N\qquad &\quad $f$(in \%)\quad
&\quad $\eps_1$ (in \%)\quad & \quad
$\eps_2$ (in \%) \quad &\cr
\+ \hfill (Sub)field  & \hfill N &\hfill $f$(in \%) &\hfill $\eps_1$ (in \%) &
$\hfill \eps_2$ (in \%) &\cr
\smallskip
\+ \hfill T&\hfill 145&\hfill  0.22&\hfill$-$2.07&\hfill$-$1.61&\cr
\smallskip
\+ \hfill A&\hfill  37&\hfill  4.80&\hfill$-$2.96&\hfill   0.94&\cr
\+ \hfill B&\hfill  40&\hfill 30.06&\hfill$-$2.22&\hfill$-$0.27&\cr
\+ \hfill C&\hfill  30&\hfill  3.85&\hfill$-$2.42&\hfill$-$4.44&\cr
\+ \hfill D&\hfill  38&\hfill  6.44&\hfill$-$0.77&\hfill$-$3.25&\cr
\smallskip
\+ \hfill 1&\hfill  15&\hfill 30.52&\hfill$-$3.13&\hfill$-$1.15&\cr
\+ \hfill 2&\hfill  22&\hfill  2.65&\hfill$-$3.98&\hfill   0.74&\cr
\+ \hfill 3&\hfill  15&\hfill 36.36&\hfill$-$3.90&\hfill$-$0.25&\cr
\+ \hfill 4&\hfill  18&\hfill 25.12&\hfill$-$3.39&\hfill   1.37&\cr
\+ \hfill 5&\hfill  15&\hfill 80.02&\hfill$-$0.55&\hfill$-$1.16&\cr
\+ \hfill 6&\hfill  15&\hfill 49.24&\hfill   2.04&\hfill$-$0.88&\cr
\+ \hfill 7&\hfill  12&\hfill 14.84&\hfill$-$1.49&\hfill$-$6.91&\cr
\+ \hfill 8&\hfill  16&\hfill  2.41&\hfill$-$2.35&\hfill$-$5.83&\cr
\+ \hfill 9&\hfill  17&\hfill 60.90&\hfill$-$0.76&\hfill$-$2.73&\cr

\medskip
\tabcap{3}{Same as Table 1, for the QSO 3C446}
\smallskip
\settabs\+ (Sub)field \quad &\qquad N\qquad &\quad $f$(in \%)\quad
&\quad $\eps_1$ (in \%)\quad & \quad
$\eps_2$ (in \%) \quad &\cr
\+ \hfill (Sub)field  & \hfill N &\hfill $f$(in \%) &\hfill $\eps_1$ (in \%) &
$\hfill \eps_2$ (in \%) &\cr
\smallskip
\+ \hfill T&\hfill 211&\hfill 20.59&\hfill$-$1.06&\hfill   0.59&\cr
\smallskip
\+ \hfill A&\hfill  64&\hfill 48.39&\hfill$-$1.06&\hfill   0.75&\cr
\+ \hfill B&\hfill  50&\hfill 47.01&\hfill$-$1.93&\hfill$-$0.18&\cr
\+ \hfill C&\hfill  55&\hfill 37.22&\hfill$-$0.19&\hfill   2.02&\cr
\+ \hfill D&\hfill  42&\hfill 67.18&\hfill$-$1.17&\hfill$-$0.58&\cr
\smallskip
\+ \hfill 1&\hfill  32&\hfill 53.31&\hfill   0.28&\hfill   1.84&\cr
\+ \hfill 2&\hfill  31&\hfill 11.36&\hfill$-$2.89&\hfill$-$2.30&\cr
\+ \hfill 3&\hfill  12&\hfill 80.48&\hfill   2.37&\hfill$-$0.60&\cr
\+ \hfill 4&\hfill  33&\hfill 64.08&\hfill   0.72&\hfill   1.35&\cr
\+ \hfill 5&\hfill  23&\hfill 66.12&\hfill$-$0.80&\hfill$-$1.55&\cr
\+ \hfill 6&\hfill  26&\hfill  0.52&\hfill$-$4.20&\hfill   4.17&\cr
\+ \hfill 7&\hfill  20&\hfill 40.92&\hfill$-$2.74&\hfill   2.29&\cr
\+ \hfill 8&\hfill  20&\hfill 98.73&\hfill   0.08&\hfill   0.39&\cr
\+ \hfill 9&\hfill  14&\hfill 69.63&\hfill$-$1.02&\hfill$-$1.87&\cr

\sec{4. Shear measurement with the ACF}
We used the ACF independently on the same images but with somewhat
different selection criteria for the galaxies than those used with BM.
As described by Van Waerbeke et al. (1997), the ACF allows a maximum
use of the signal which increases significantly the signal-to-noise
ratio on the measurement of the local shear. Ideally, all the pixels
of the images could be used, but in practice we experienced that a
better signal-to-noise ratio can be achieved if the ACF is computed on
restricted areas centered around each galaxy, previously detected by a
standard technique.  The main reason for this is that, as suggested
from the deep HDF image, even at extremely faint magnitude, a large
fraction of the CCD frames remains free from very faint sources,
showing that the faint-end slope of the galaxy number counts decrease
substantially and cannot provide a complete coverage of the whole
field with faint galaxies, down to the confusion limit. This fact
limits the efficiency of the ACF in `empty regions' of the CCD since
only a low signal is coming from backgound sources hidden in the
noise. In addition, as we shall see further below, the data
acquisition and the pre-processing procedures (shift-and-add) as well
as the electronic boards of the CCD camera generate correlated
residual noise which perturb the measurement of the ellipticities of
very faint galaxy images. We therefore decided to optimize the
algorithm by using first SExtractor for source detection, and then by
applying the ACF around those sources only. The efficiency of this
method has been discussed already by Van Waerbeke \& Mellier (1997).

In this Section we focus on the results from the ACF method and the
comparison with the SExtractor measure using exactly the same
galaxies. Futhermore we will compare these results with the previous
BM analysis.

For each field, a new SExtractor catalogue has been built using
standard but different selection criteria than the previous
analysis. Typically the convolution filter is larger and does not
permit the detection of faint peaked objects.

The catalogues contain $N_{\rm g}=145$ galaxies for PKS1508, $N_{\rm
g}=129$ for PKS0135 and $N_{\rm g}=181$ for 3C446.  Only objects with
FLAG=0 are keeped, to be sure to eliminate all the possible external
source of distortion, and the magnitude ranges are $[22,28]$,
$[21.3,27.3]$, and $[22.3,28.3]$. These restrictive conditions explain
why the number of galaxies is slightly smaller than in the BM
catalogues.  As a first step, the mean ellipticity was calculated in
the whole fields. Since the seeing on these fields ranges from
$0\arcsecf66$ to $0\arcsecf76$, this corresponds to a gaussian PSF of
size 5 to 6 pixels in the ACF space. This means that, ideally, we have
to use an isophotal annular filter of mean diameter equal to 6 pixels
to compute the shape parameters of the ACF. In Figs.\ts 8 through 10,
the ACFs for the three QSO fields are shown.  These figures reveal the
fact that the ACF's center is polluted by an instrumental small-scale
pixel-pixel correlation probably due to the reasons discussed
above. To avoid this center we used an annular filter with 10 pixels
diameter, larger than the ideal filter size.  The inconvenience is
that at larger distance from the ACF's center, the S/N is lower. The
ellipticities are computed in different magnitude bins.  The results
are shown in Figs.\ts 5 through 7. The shear detection in PKS1508 is
robust and confirms the results of Sect.\ts 3. The other two field do
not show a significant distortion, although a significant shear is
detected in the field of PKS0135 if only the brighter galaxies are
used. The agreement between SExtractor and ACF is very good.  However,
it seems that at magnitudes fainter than $27$ the ACF measures a
larger signal. In fact, because for the faintest galaxies, the
instrumental structure in the ACF (Figs.\ts 8 through 10) becomes
important, the measurement is not accurate. To see how this
intrumental effect alters the ellipticity measurement, Fig.\ts 11
shows the same ellipticity as Fig.\ts 5, but using an annular filter
closer to the ACF center, with a diameter of $6$ pixels. The
anormalously high value of the ellipticity is clear, and is due to the
fact that the instrumental distortion is in the same direction as the
ellipticity orientation. Only a large diameter for the filter and/or
the use of the brighter galaxies only allows to avoid this problem.

The measurements were performed in the whole field T and in the subfields
ABCD using a more restrictive condition on the magnitude of the
objects. The objects are brighter than a threshold magnitude
$m_{\rm lim}$. The results are shown on Tables 4 through 6. The number of
galaxies is significantly smaller than in the Tables 1, 2 \& 3 for the
BM analysis because of our more restrictive detection conditions
here. While the ACF and SExtractor results are consistent, it is
difficult to compare them with BM in the ABCD fields because the
galaxy samples are different. However, in the T fields, the number of
galaxies is large and the results are consistent.  The detection of
the shear in PKS1508 is confirmed. It was impossible to performed the
calculations in the subfields 1-9 because of the very small number of
galaxies.  We note that some subfields (D in PKS1508 and C in PKS0135)
show a large value of the shear, with a significance level
of $4.2 \sigma$\fonote{Here, the significance level is determined from
the random intrinsic ellipticity of the galaxies, which is very close
to the rms ellipticity of the observed images.} and $2.8 \sigma$,
respectively, with the ACF.  An 
interesting point is the $2.9 \sigma$ detection of an homogeneous shear
in the whole field of PKS0135, if only the brightest galaxies
($m_{\rm lim}=23.5$) are kept.

Since it is difficult to compare with the BM results, because the
selection criteria are not {\it exactly} the same, we decided to
perform an additional analysis which consists of using the three
methods on a set of galaxies which are present in the three
catalogues. For each field, the resulting catalogue corresponds to
highly conservative selection criteria. The number of objects are,
respectively, 93, 73 and 128 for the PKS1508, PKS0135, and 3C446
fields. The BM ellipticities are recalibrated with SExtractor
ellipticities. We find that the calibration constants $C$ are changed
compared to Sect.\ts 3, which demonstrates that this calibration
depends on the magnitude of the galaxies, which is not surprising. The
results are shown in Tables 7, 8, 9. The agreement between the methods
is remarkable, and confirms the detection of a cosmic shear in
PKS1508. The polar coordinates of the shear $(|\epsilon |,\theta)$ are also
displayed for easier visualisation of the geometry of the distortion.
The small excess in the intensity measured by the ACF compared to
SExtractor is probably due to the fact that SExtractor measures the
image brightness moments only from those pixels above a given flux
threshold.  This confirms the results obtained from simulated images
(Van Waerbeke 1997).

\xxfigure{5}{In the left panel, the value of the mean ellipticity of the shear
in PKS0135 as a function of the upper bound value $m_{\rm lim}$ of the
magnitude interval is plotted. The right panel shows the orientation
angle of the mean ellipticity. 
The solid curves correspond to the ACF, and the dashed curves to
the SExtractor results. The errors bars are the $1\sigma$ intervals
due to the intrinsic ellipticities of the galaxies; note that the decrease of the
size of the error bars due to the increase in galaxy number with fainter
magnitude is partly compensated by increases ellipticities of galaxy images, which
is mainly due to the increasing noise.
}{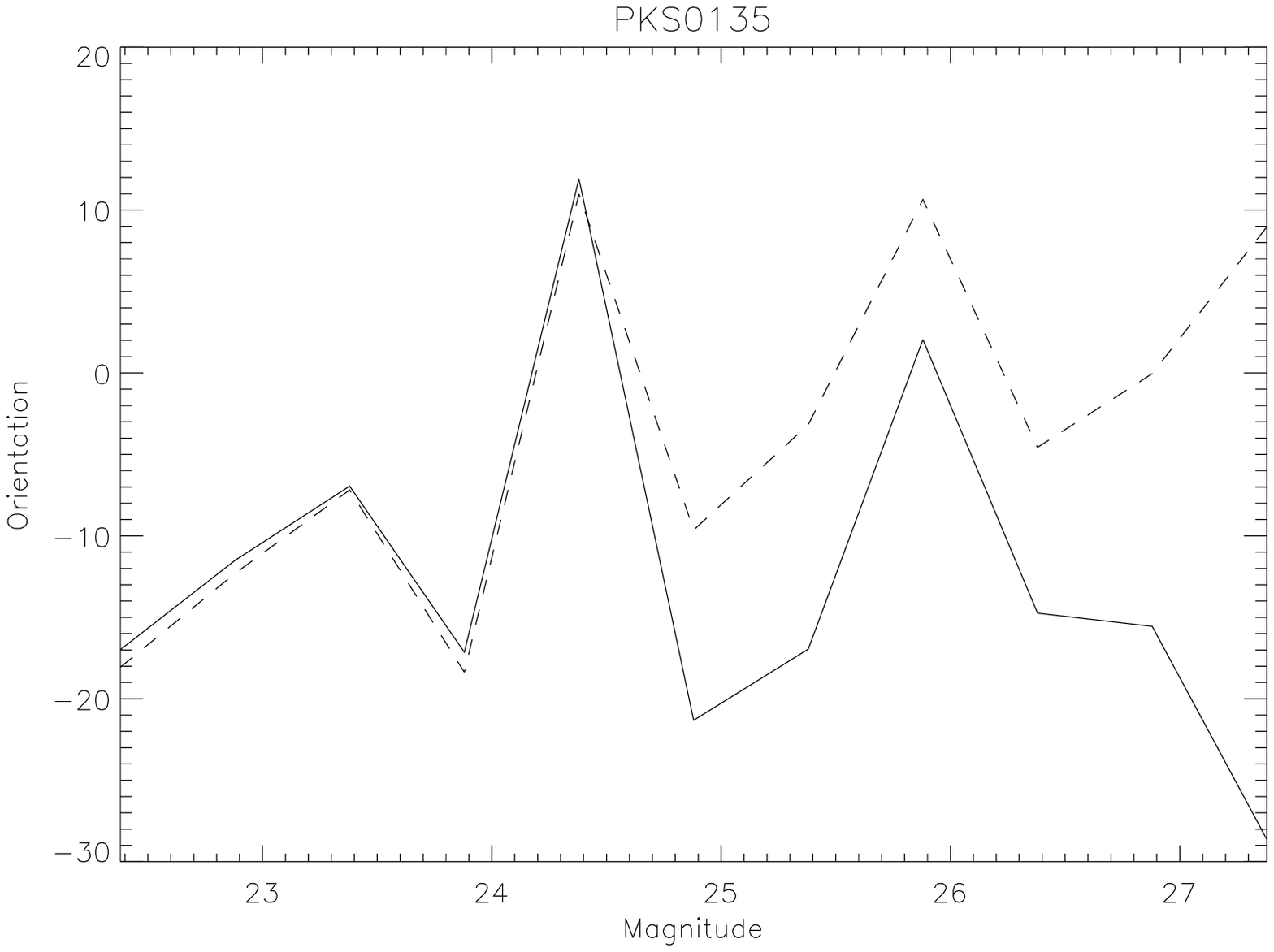}{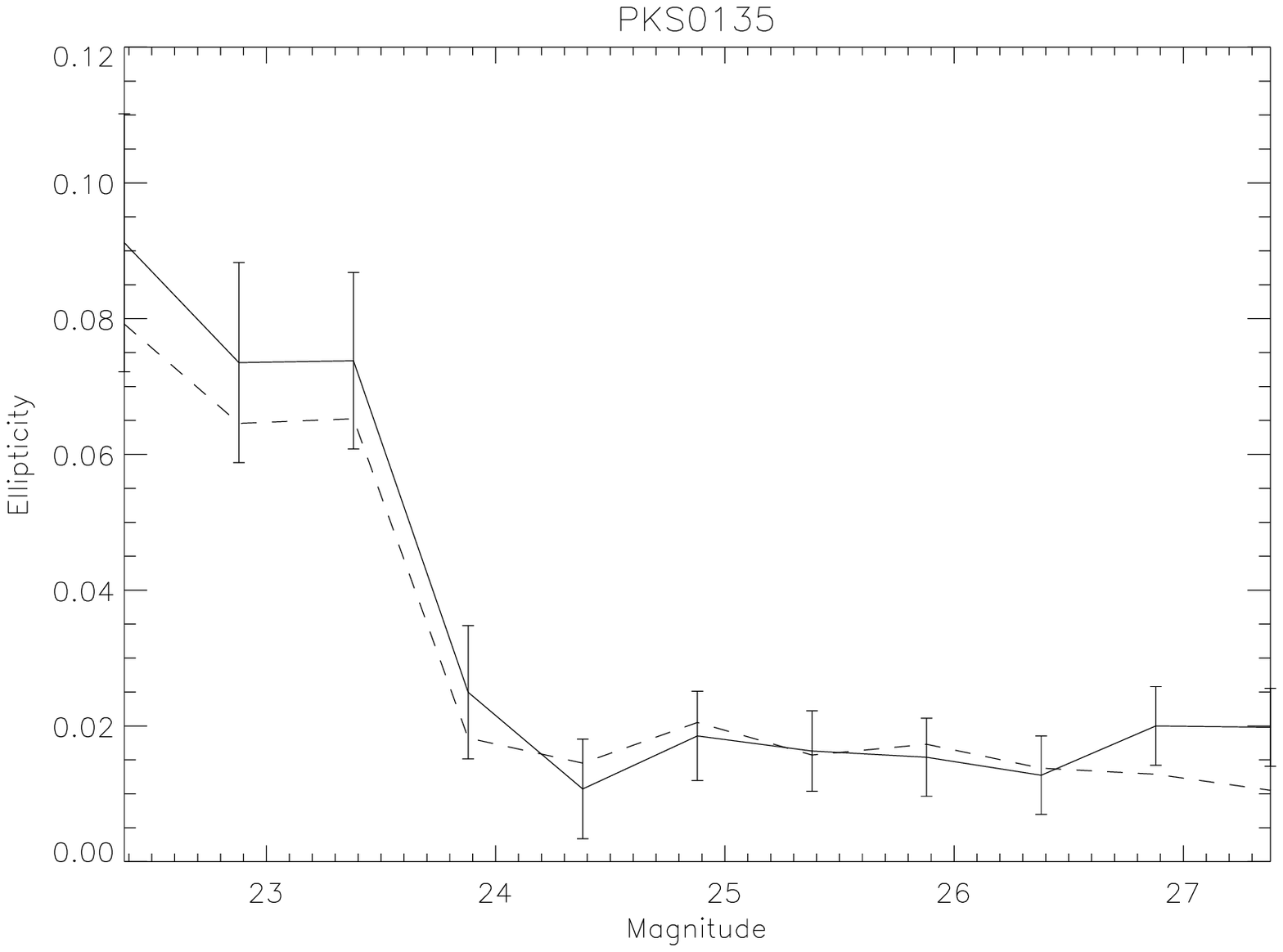}{8}{8}

\xxfigure{6}{Same as Fig.\ts 5 for the PKS1508 field
}{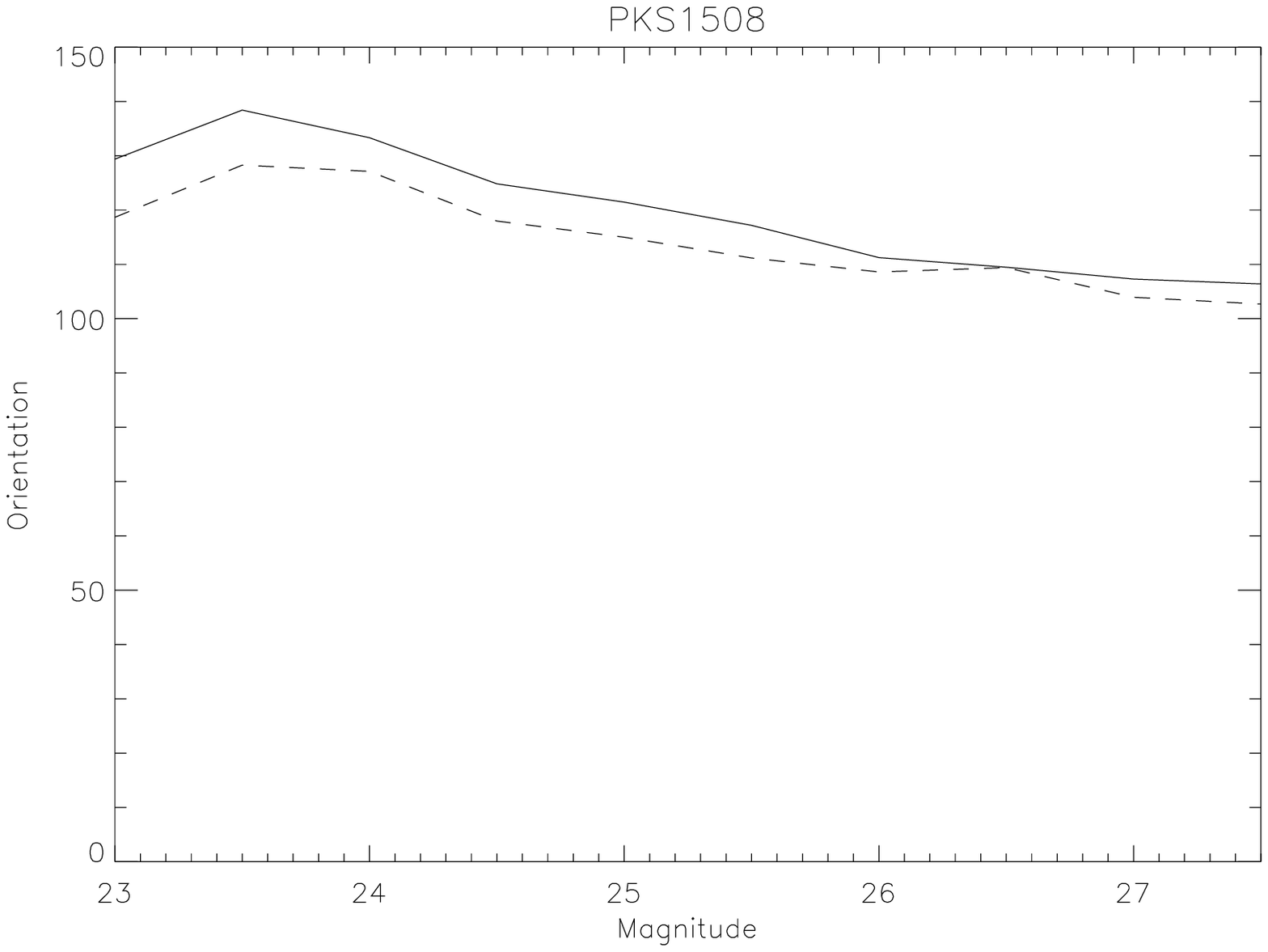}{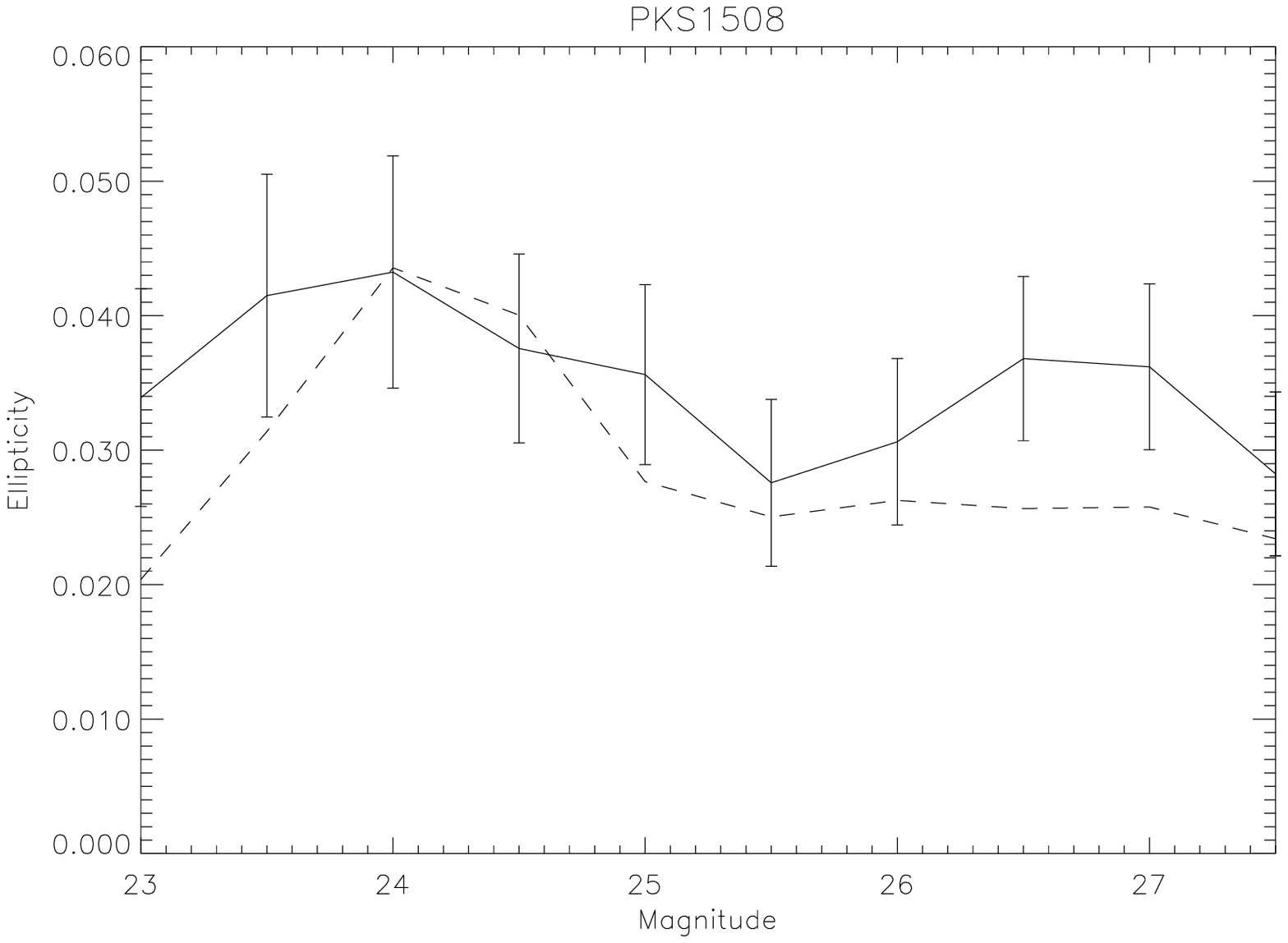}{8}{8}

\xxfigure{7}{Same as Fig.\ts 5 for the 3C446 field
}{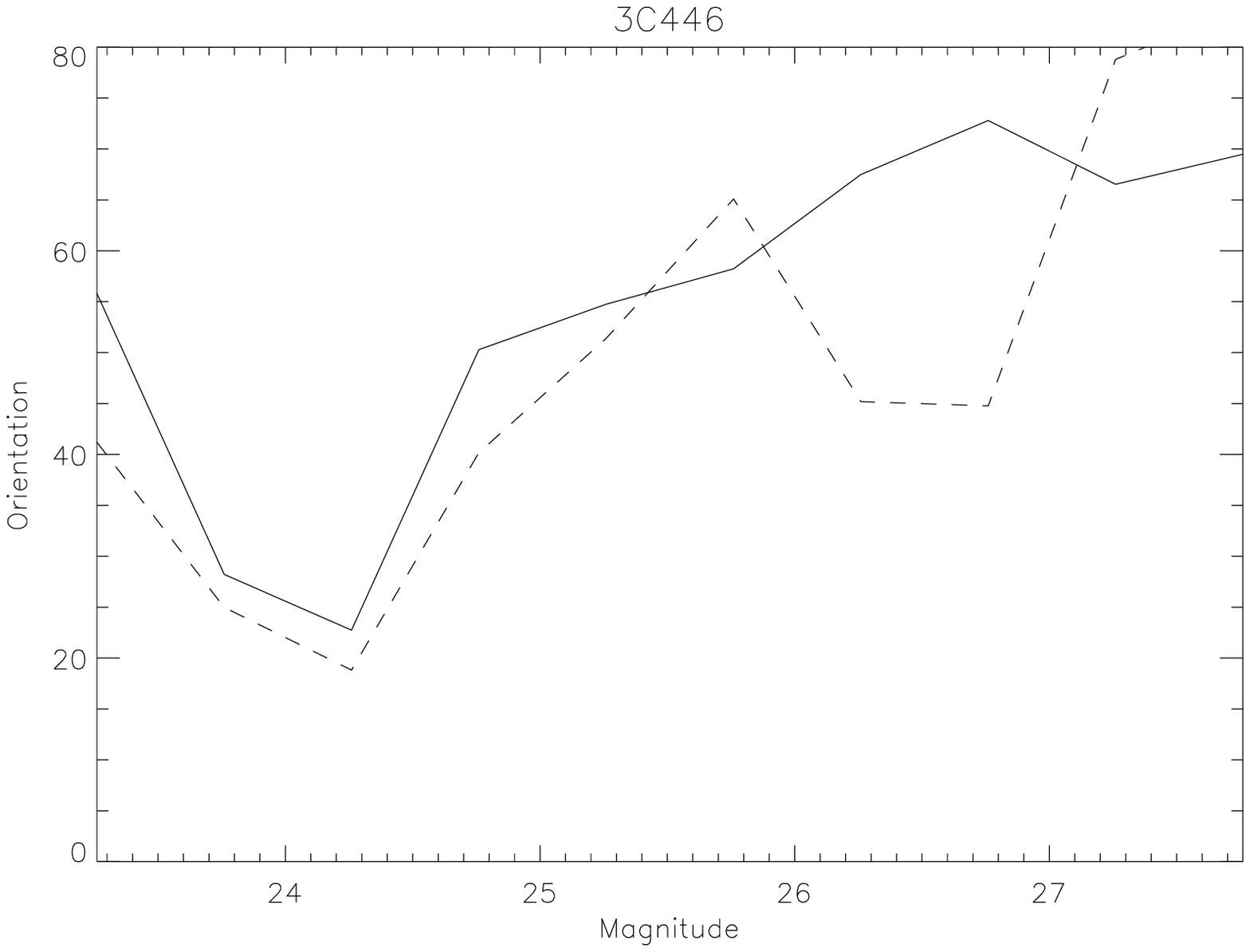}{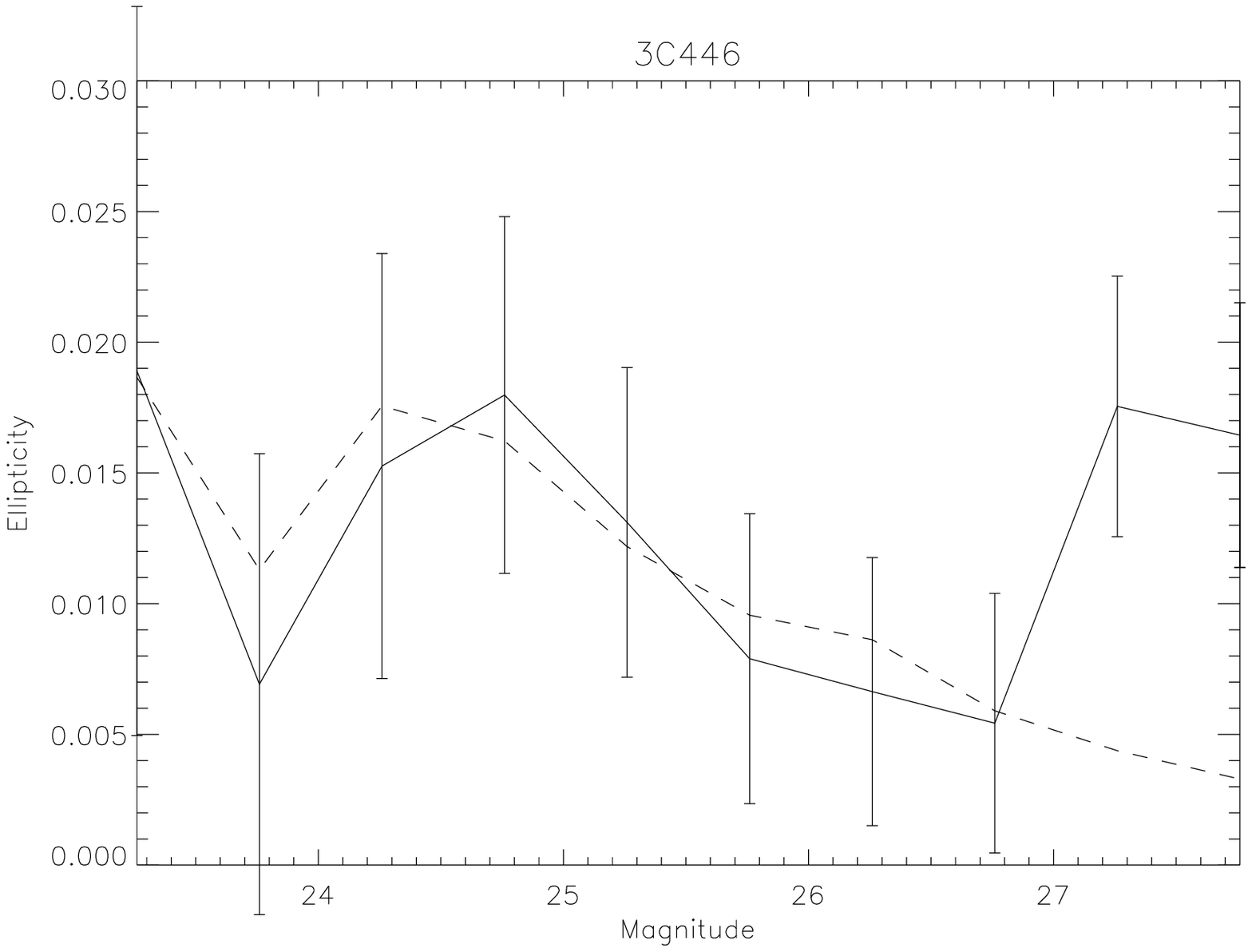}{8}{8}

\xxfigure{8}{The left panel shows the ACF of the faint galaxies ($m\in [21.3,27.3]$)
 in the field PKS0135. The central structure is probably due
 to correlated pixel-to-pixel noise due to CCD
read-out and data reduction processes. On the right, the
 contour levels show this internal structure and the regularity of the
 ACF's isophotes.  }{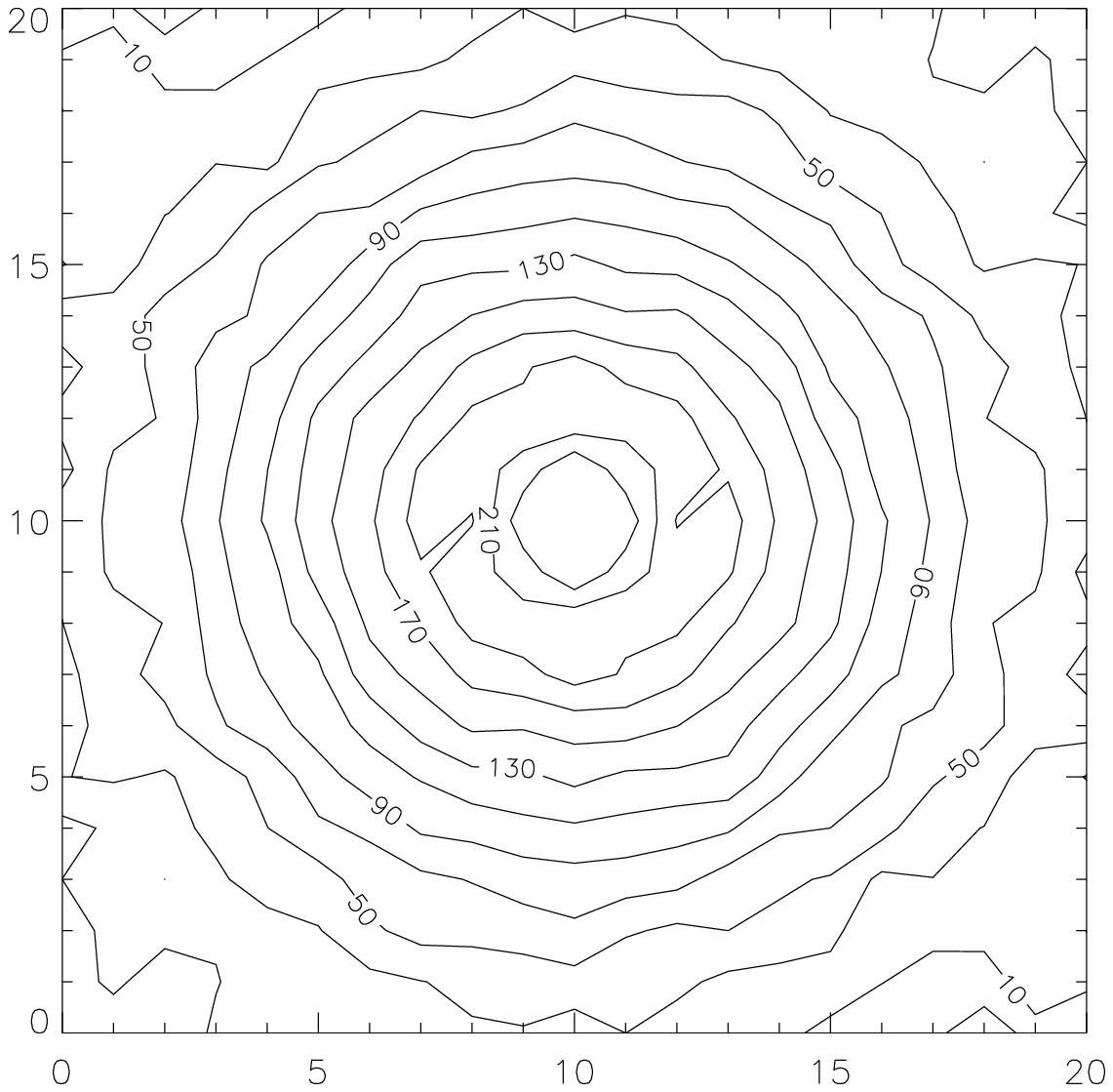}{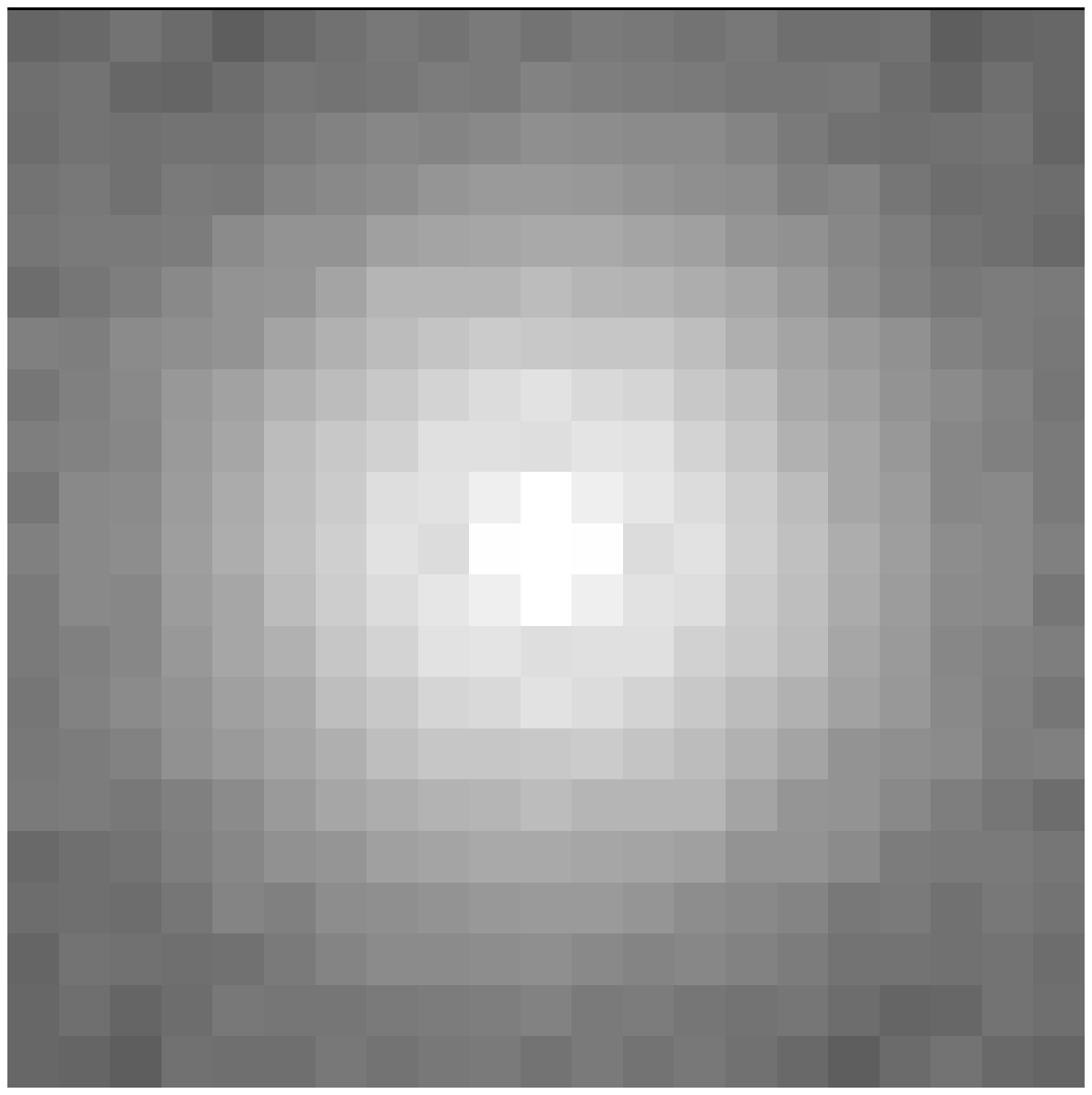}{6}{9}

\xxfigure{9}{Same as Fig. \ts 8 for the PKS1508 field. The magnitude
range is $m\in [22,28]$.
}{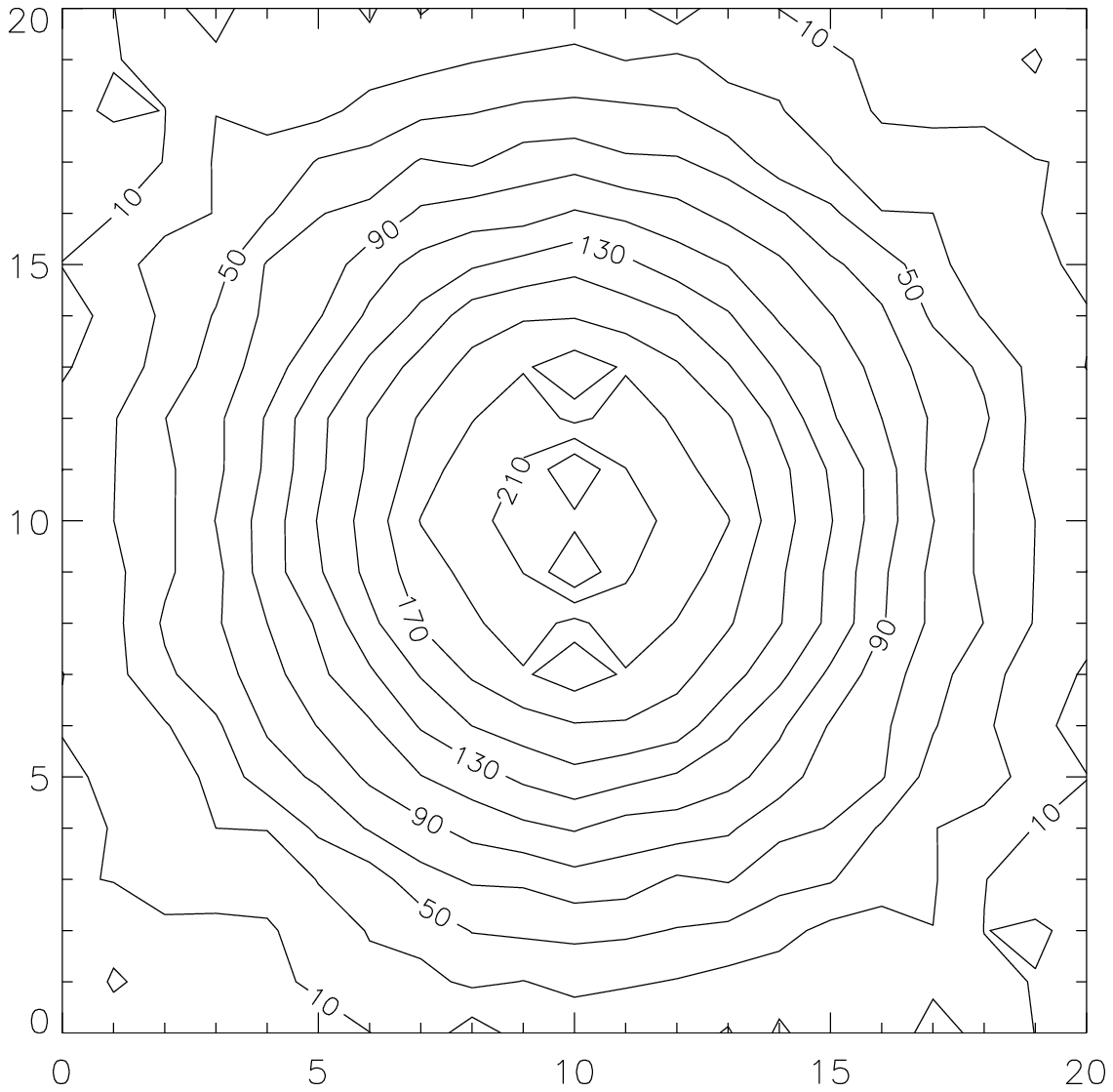}{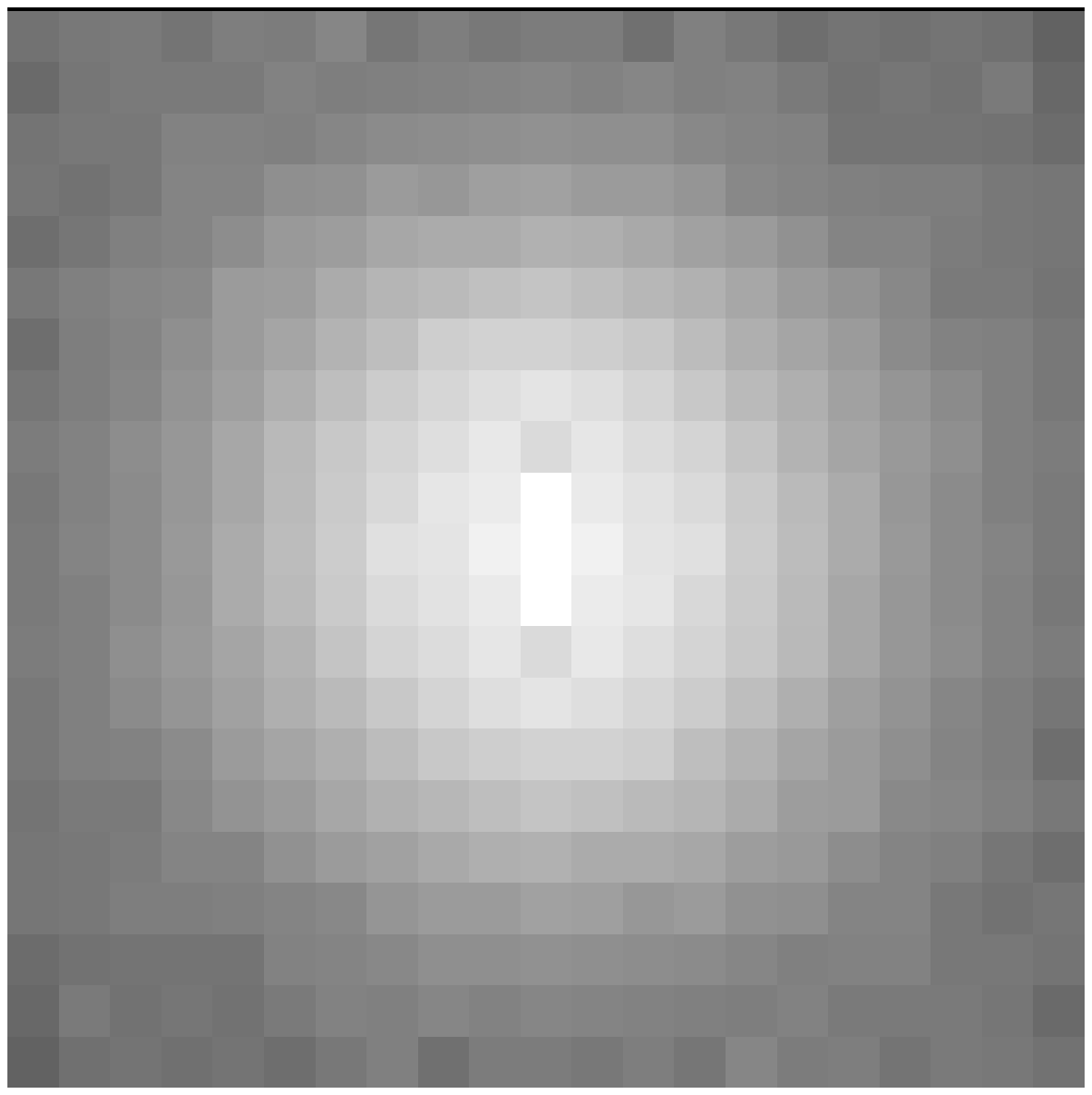}{6}{9}

\xxfigure{10}{Same as Fig. \ts 8 for the 3C446 field. The magnitude range is $m\in [22.3,28.3]$.
}{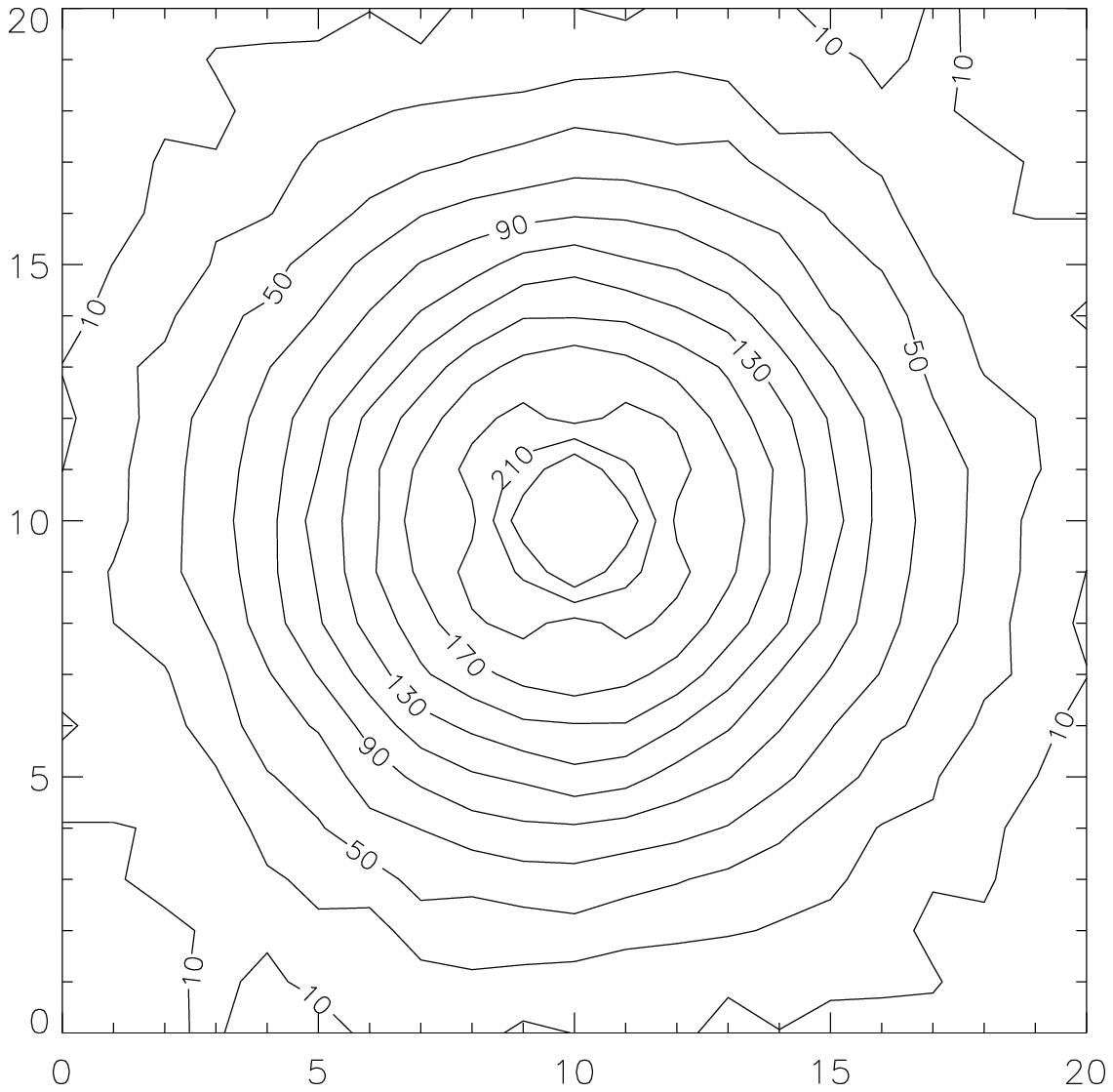}{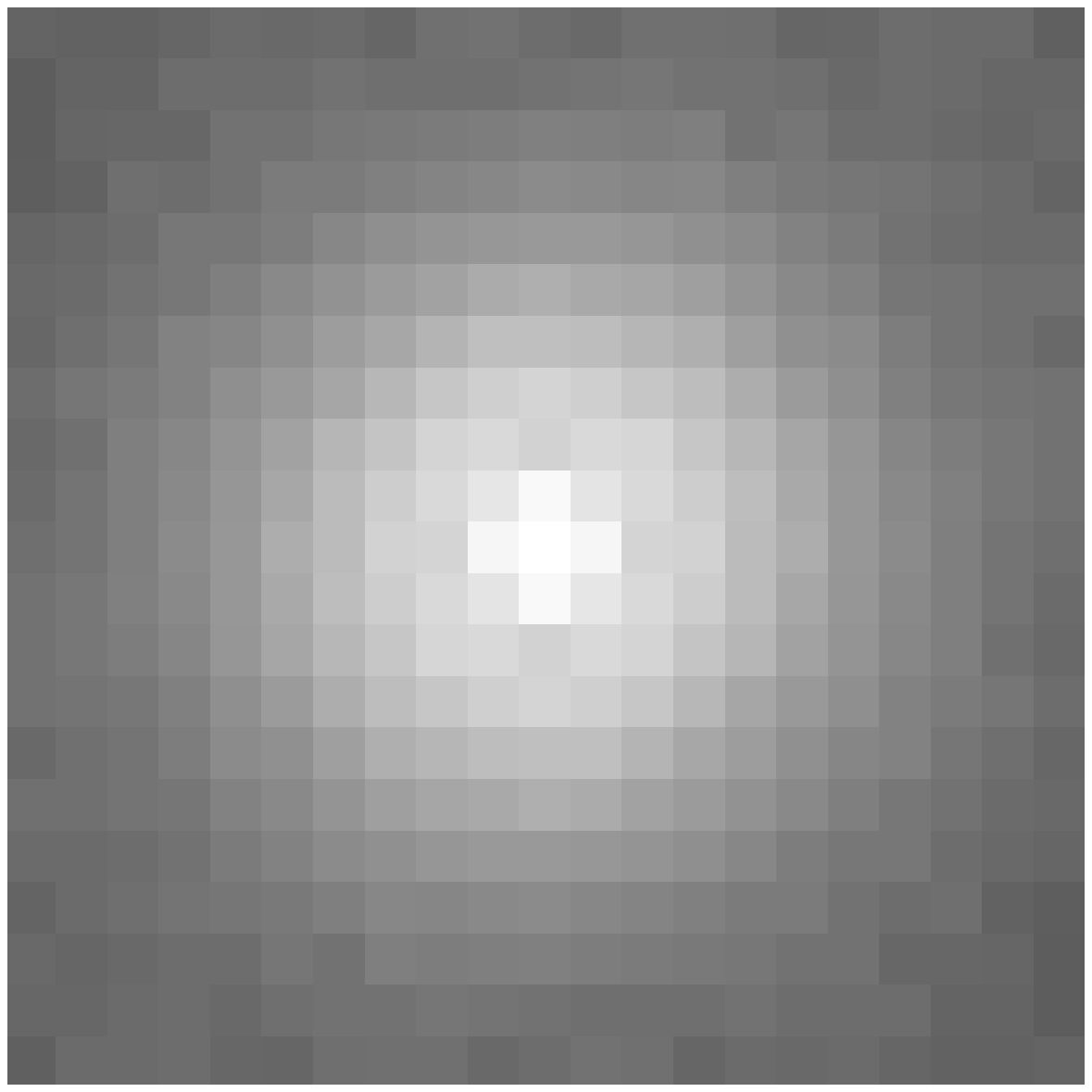}{6}{9}

\xfigure{11}{Same as Fig.\ts 6, but with a smaller annular isophotal
filter. The larger value of the shear compared to Fig.\ts 6 in the ACF
method is due to the internal structure showed on Fig.\ts 9
}{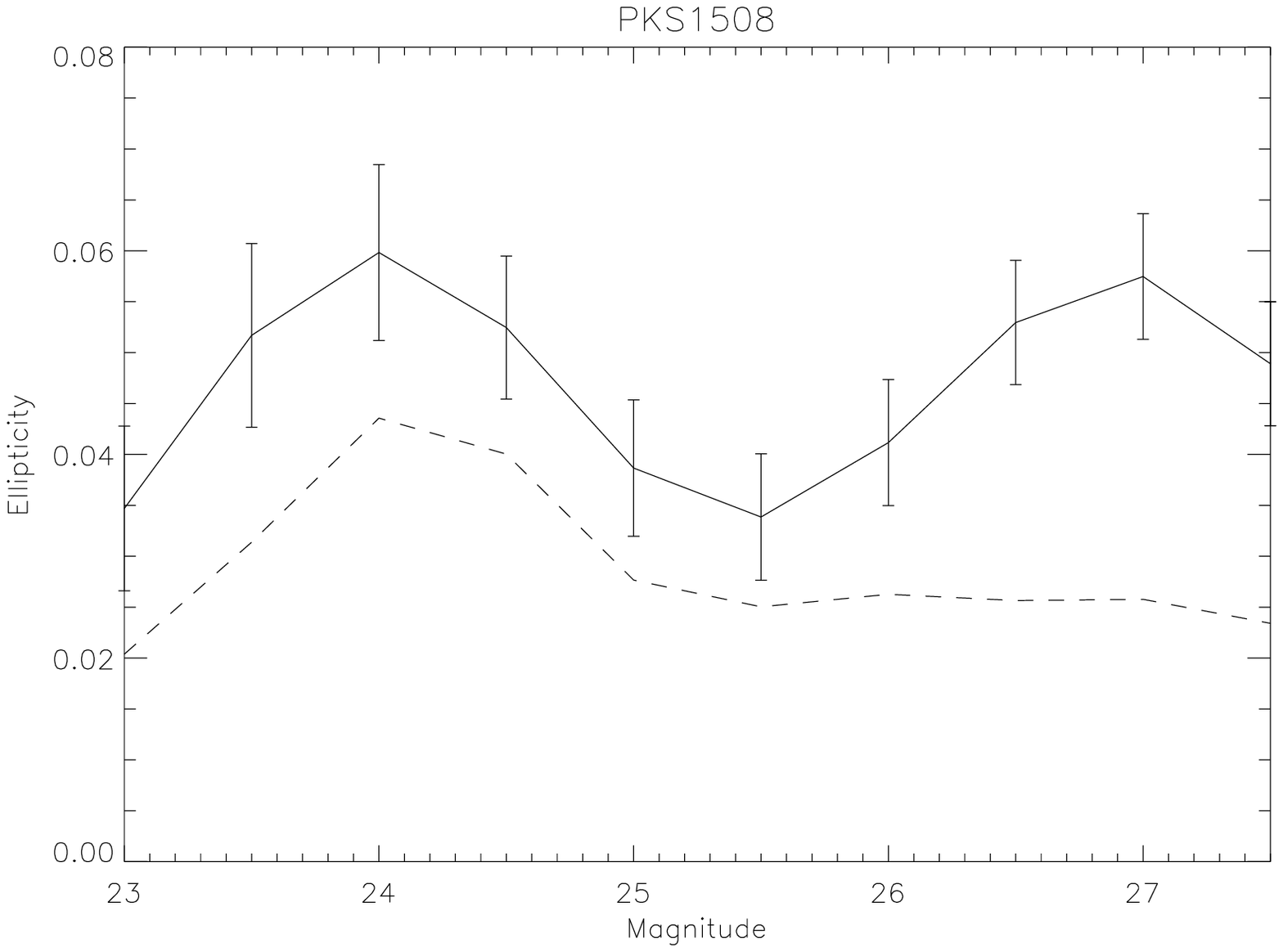}{10}

\vfill\eject

\medskip
\tabcap{4}{For the QSO 0135 and the TABCD (sub)fields as described in the
text, the number $N$ of 
objects, and the measured mean
ACF ellipticity is compared to the S-ellipticity. 
The upper magnitude limit is $m_{\rm lim}=25.3$}
\smallskip
\settabs\+ (Sub)field \quad &\qquad N\qquad
&\quad $\eps_1^{\rm ACF}$ \quad & \quad
$\eps_2^{\rm ACF}$  \quad & \quad $\eps_1^{\rm S}$ \quad & \quad
$\eps_2^{\rm S}$ \quad &\cr
\+ \hfill (Sub)field  & \hfill N &\hfill $\eps_1^{\rm ACF}$  &
$\hfill \eps_2^{\rm ACF}$  & \hfill $\eps_1^{\rm S}$  &$\hfill
\eps_2^{\rm S}$ &\cr
\smallskip
\+ \hfill T&\hfill  81&\hfill    1.35&\hfill     $-$0.91&\hfill 1.55&\hfill $-$0.17&\cr
\smallskip
\+ \hfill A&\hfill   19&\hfill    0.98&\hfill  0.56&\hfill  2.94&\hfill $-$0.02&\cr
\+ \hfill B&\hfill   23&\hfill    $-$0.52&\hfill  $-$0.53&\hfill  $-$0.81&\hfill 0.14&\cr
\+ \hfill C&\hfill   24&\hfill    5.16&\hfill  $-$2.98&\hfill  3.41&\hfill $-$1.81&\cr
\+ \hfill D&\hfill   15&\hfill    $-$1.34&\hfill  0.14&\hfill  0.69&\hfill 1.70&\cr

\vfill\eject

\medskip
\tabcap{5}{For the QSO 1508 and the TABCD (sub)fields as described in the
text, the number $N$ of 
objects, and the measured mean
ACF-ellipticity is compared to the S-ellipticity. 
The upper magnitude limit is $m_{\rm lim}=26.$}
\smallskip
\settabs\+ (Sub)field \quad &\qquad N\qquad
&\quad $\eps_1^{\rm ACF}$ \quad & \quad
$\eps_2^{\rm ACF}$  \quad & \quad $\eps_1^{\rm S}$ \quad & \quad
$\eps_2^{\rm S}$ \quad &\cr
\+ \hfill (Sub)field  & \hfill N &\hfill $\eps_1^{\rm ACF}$  &
$\hfill \eps_2^{\rm ACF}$  & \hfill $\eps_1^{\rm S}$  &$\hfill
\eps_2^{\rm S}$ &\cr
\smallskip
\+ \hfill T&\hfill  118&\hfill    $-$2.26&\hfill     $-$2.03&\hfill $-$2.09&\hfill $-$1.59&\cr
\smallskip
\+ \hfill A&\hfill   33&\hfill    $-$3.80&\hfill  0.08&\hfill  $-$3.38&\hfill 0.09&\cr
\+ \hfill B&\hfill   28&\hfill    $-$3.00&\hfill  $-$0.56&\hfill  $-$1.84&\hfill $-$1.15&\cr
\+ \hfill C&\hfill   25&\hfill    $-$2.46&\hfill  0.20&\hfill  $-$1.94&\hfill $-$0.51&\cr
\+ \hfill D&\hfill   32&\hfill    0.61&\hfill  $-$7.76&\hfill  $-$1.14&\hfill $-$4.61&\cr
\medskip

\tabcap{6}{For the QSO 3C446 and the TABCD (sub)fields as described in the
text, the number $N$ of 
objects, and the measured mean
ACF-ellipticity is compared to the S-ellipticity. 
The upper magnitude limit is $m_{\rm lim}=26.2$}
\smallskip
\settabs\+ (Sub)field \quad &\qquad N\qquad
&\quad $\eps_1^{\rm ACF}$ \quad & \quad
$\eps_2^{\rm ACF}$  \quad & \quad $\eps_1^{\rm S}$ \quad & \quad
$\eps_2^{\rm S}$ \quad &\cr
\+ \hfill (Sub)field  & \hfill N &\hfill $\eps_1^{\rm ACF}$  &
$\hfill \eps_2^{\rm ACF}$  & \hfill $\eps_1^{\rm S}$  &$\hfill
\eps_2^{\rm S}$ &\cr
\smallskip
\+ \hfill T&\hfill  140&\hfill    $-$0.46&\hfill     0.48&\hfill $-$0.06&\hfill 0.86&\cr
\smallskip
\+ \hfill A&\hfill   43&\hfill    0.82&\hfill  2.25&\hfill  0.34&\hfill 1.91&\cr
\+ \hfill B&\hfill   27&\hfill    $-$1.63&\hfill  $-$1.47&\hfill  2.24&\hfill $-$0.52&\cr
\+ \hfill C&\hfill   34&\hfill    $-$0.63&\hfill  2.71&\hfill  $-$0.38&\hfill 2.91&\cr
\+ \hfill D&\hfill   36&\hfill    $-$0.92&\hfill  $-$2.84&\hfill  $-$1.78&\hfill $-$1.20&\cr

\bigskip

\tabcap{7}{Comparison of the 3 methods used in this paper (ACF, BM,
and SExtractor) for PKS0135 using the whole image. A total of 73
objects are kept, only those which are common to the 3 catalogues and
with FLAG=0. The limiting magnitude here is $m_{\rm lim}=25.3$. The quantity
$(|\epsilon |,\theta)$ is the mean ellipticity and mean orientation.}
\smallskip
\settabs\+ Methods \quad &\qquad $\eps_1$\qquad
&\quad $\eps_2$ \quad & \quad
$|\epsilon |$\quad & \quad $\theta$ \quad &\cr
\+ \hfill Methods & \hfill $\eps_1$ &\hfill $\eps_2$  &
$\hfill |\epsilon |$  & \hfill $\theta$  &\cr
\smallskip
\+ \hfill ACF&\hfill  1.43&\hfill   $-$0.89&\hfill  1.68&\hfill $-$16&\cr
\+ \hfill BM&\hfill   1.37&\hfill   0.42&\hfill  1.43&\hfill 8&\cr
\+ \hfill Sex&\hfill  1.26&\hfill   $-$0.68&\hfill  1.43&\hfill $-$14&\cr

\vfill\eject

\tabcap{8}{Same as Table 7, for PKS1508, with 93 objects, and $m_{\rm lim}=26$.}
\smallskip
\settabs\+ Methods \quad &\qquad $\eps_1$\qquad
&\quad $\eps_2$ \quad & \quad
$|\epsilon |$  \quad & \quad $\theta$ \quad &\cr
\+ \hfill Methods & \hfill $\eps_1$ &\hfill $\eps_2$  &
$\hfill |\epsilon |$  & \hfill $\theta$  &\cr
\smallskip
\+ \hfill ACF&\hfill  $-$2.28&\hfill   $-$3.62&\hfill  4.28&\hfill 119&\cr
\+ \hfill BM&\hfill   $-$2.83&\hfill   $-$2.53&\hfill  3.80&\hfill 111&\cr
\+ \hfill Sex&\hfill  $-$2.58&\hfill   $-$2.78&\hfill  3.80&\hfill 114&\cr

\medskip

\tabcap{9}{Same as Table 7 for 3C446 128 objects, and $m_{\rm lim}=26.3$.}
\smallskip
\settabs\+ Methods \quad &\qquad $\eps_1$\qquad
&\quad $\eps_2$ \quad & \quad
$|\epsilon |$  \quad & \quad $\theta$ \quad &\cr
\+ \hfill Methods & \hfill $\eps_1$ &\hfill $\eps_2$  &
$\hfill |\epsilon |$  & \hfill $\theta$  &\cr
\smallskip
\+ \hfill ACF&\hfill  $-$1.30&\hfill   1.15&\hfill  1.74&\hfill 69&\cr
\+ \hfill BM&\hfill   $-$1.11&\hfill   0.97&\hfill  1.47&\hfill 69&\cr
\+ \hfill Sex&\hfill  $-$1.00&\hfill   1.09&\hfill  1.47&\hfill 66&\cr

\sec{5. The two-point correlation function of the galaxy
ellipticities}
We have analyzed the mean shear present in the three QSO fields
in sections 3 and 4. The fields were subdivided and the mean shear in 
sub-fields was compared, and its significance tested by randomizing
the ellipticity position angles to generate mock data. In this
section we analyze the data in a somewhat complementary way by
computing the ellipticity auto-correlation function defined as
$$
C_{\eps\eps}(\theta) =  \langle {\bf \epsilon} (\vc\vp) \
{\bf\epsilon}^*(\vc\vp+\vc\theta)\rangle_{\vc\vp}  , 
\eqno(2)
$$
where the angular brackets denote ensemble averaging. For a finite set
of data it is evaluated by considering all pairs available with
angular separation in a small interval around
$\theta=\abs{\vc\theta}$. Note that the correlation function $C_{pp}$
defined in Jain \& Seljak (1996) uses an ellipticity estimate which
for small ellipticities is twice as large as $\eps$, and so
$C_{pp}(\theta)=4 C_{\eps\eps}(\theta)$; for notational consistency,
we shall use $C_{\eps\eps}$ henceforth.  While $C_{\eps\eps}$ does not
retain any information about the location of the galaxies, it
highlights the relative contributions to the mean shear from
correlations on different angular scales. It is also a convenient
statistic to compare with theoretical predictions.  As discussed
below, ideally one would like at least 10 well separated fields to get
a sensible measurement of $C_{\eps\eps}$ for comparison with
theoretical predictions. For just one field, it is more useful as a
complementary check of the robustness of the signal and a test of
approximate quantitative agreement with predictions.

We typically have about 150 galaxies in fields of $2'$ on a side.  We
compute $C_{\eps\eps}$ by binning pairs into 5 bins logarithmically
spaced in $\theta$. Each bin, except for the first, has over 2000
pairs of BM ellipticities which we average over to compute
$C_{\eps\eps}$ as defined in Eq.\ts(2).  We found that for the fields
of QSO 3C446 and PKS0135, the data did not provide a non-zero
$C_{\eps\eps}$ that was robust to resampling and small variation of
the bin size.  For the field of PKS1508 however, we measure a robust,
non-zero value of $C_{\eps\eps}$ for 3 of the 5 angular bins.

\xfigure{12}{$C_{\eps\eps}(\theta)$ computed using the Bonnet-Mellier
ellipticities for the field of PKS1508 (filled dots), plotted
vs. $\theta$ in arcminutes. The solid error bars are the 1-$\sigma$
bootstrap resampling deviations. The circles and the dashed error bars
(slightly displaced in $\theta$ for clarity) show
$C_{\eps\eps}(\theta)$ computed for simulated data obtained by
randomizing the position angles of the measured galaxy ellipticites,
and the corresponding 1-$\sigma$ deviations for 10,000 such data
sets. The solid curve is the fit $C_{\eps\eps}=a/(\theta+1')$ to the
measured data (see text for details).}{fig88.tps}{10}

The results for PKS1508 are shown in Figure 12 which shows
$C_{\eps\eps}(\theta)$ vs. $\theta$ in arcminutes. The solid error bars are
the 1-$\sigma$ deviations obtained from bootstrap resampling of the
data. A closer examination of the distribution of the bootstrap
resampled $C_{\eps\eps}$'s showed a positively skewed distribution.
The true error bars are therefore shifted slightly upward
from what is shown, thus providing a firmer lower limit. 
The detected value of $C_{\eps\eps}$ is above the
1-$\sigma$ level for 3 of the bins, and for the central bin centered
on $0\arcminf 62$ it is above the 2-$\sigma$ level.

The robustness of the detection of a non-zero $C_{\eps\eps}$ can be further
tested by randomizing the position angles of the measured galaxy
ellipticites as before. 10,000 such simulated data sets were used to
obtain the results shown as circles and dashed error bars in
Figure 88. The result
as expected is a zero average value of $C_{\eps\eps}$; more useful are the
size of the error bars which provide an alternative estimate of the
significance level of the values plotted in the left panel. We
found that less than $0.4\%$ of the simulated data yielded a value
of $C_{\eps\eps}$ greater than or equal to the value measured from the
data at $\theta=0\arcminf 62$. The (dashed) error bars obtained from
randomizing the orientations of galaxies are more relevant for the
statistical significance for the deviation of $C_{\eps\eps}$ from that
expected for a random ellipticity distribution.

Theoretical cosmological models, normalized empirically (to the
abundance of rich galaxy clusters, following White et al. 1993) predict
a value of $C_{\eps\eps}$ for $\theta=1'$ of about $5\times 10^{-4}$
for reasonable values of $\Omega$ and $\Lambda$ (Jain \& Seljak
1996).  COBE-normalized models tend to
predict higher $C_{\eps\eps}$, so that an upper limit for
$C_{\eps\eps}(\theta=1')\simeq 1.2 \times 10^{-3}$ is
reasonable. Larger values would severely conflict with data from
large-scale structure and other studies. For CDM-like models,
$C_{\eps\eps}$ varies as $1/\theta$ for $\theta>1'$ and approaches a
constant for $\theta$ much smaller than $1'$.

With this theoretical bias in mind, we fitted a function of the 
form $a/(1'+\theta)$, with the amplitude $a$ 
fitted from the measured $C_{\eps\eps}$. The measured values in 
the 5 bins were combined, inversely
weighted by the variance, to get a best fit value for $a$. 
We obtained the result $a=0.91\times 10^{-4}$, and found that 
fewer than $0.4\%$ of the randomized samples yielded a value of $a$
larger than this value; this fit is also plotted in Fig.\ts 12. It
should be noted that this functional form does not fit very well to
the measured data points; by using a somewhat more complicated shape
of the fit function, a higher significance can be obtained, but we
have not tried this fine-tuning.

In order to compare the amplitude of the measured $C_{\eps\eps}$ with
the theoretical results we must correct the Bonnet-Mellier
ellipticities to estimate the true $C_{\eps\eps}$. Since the
correction factor is approximately 5, we obtain $a_{\rm
corrected}\simeq 0.002$, or $C_{\eps\eps}(1')\simeq 10^{-3}$ which is
close to the upper limit of the theoretical prediction. While a single
field of view $2'$ on a side is too small to draw quantitative
conclusions from, it is interesting that the measured values of
$C_{\eps\eps}$ are within the range of theoretical predictions.

\sec{6. Discussion and conclusions}

Using various methods, we have shown that the shear found in FMDBK
around at least one high-redshift QSO is statistically highly
significant. Whereas the significance of the shear over the full
fields of PKS0135$-$247 and 3C446 is not very high, the
probability to find a shear in excess of that seen in PKS1508$-$05
from a randomly oriented sample of galaxy images is below $0.4\%$, as
demonstrated by several methods. The main reason for this high level of
significance is the remarkable circularity of
the PSF as measured from isolated stellar objects in the SUSI fields. 
Moreover the residual anisotropy of the PSF in the field of
PKS1508$-$05, of order 1\%, was shown to be oriented perpendicular
to the mean ellipticity of galaxy images and therefore cannot cause
the observed statistical alignment.  
The shear in the field of QSO Q1622+236,
where FMDBK also found a significant signal, was not considered in the
present paper as it was observed with a different telescope.

For the interpretation of the statistically significant shear in the
field of PKS1508$-$05 one has to consider the field selection employed
by FMDBK. Their main motivation was to test the hypothesis put
foreward by Bartelmann \& Schneider (1992) that the observed
associations of foreground galaxies with high-redshift QSOs on
arcminute scales is due to
lensing by the large-scale structure in which the galaxies are
embedded, and thus overdense in regions of high magnification by
lensing. Therefore, FMDBK have selected several high-redshift QSOs
with large flux in both optical and radio wavebands. If the Bartelmann
\& Schneider hypothesis is true, then the lines-of-sight selected by
FMDBK are biased for the presence of a lensing effect. 

We therefore discuss three alternative interpretations of the detected
signal: 

(1) The first is to assume that the QSO-galaxy associations on arcminute
scales is a statistical fluke, or that it is unrelated to gravitational
lensing. Given the increasing evidence for such correlations (e.g.,
Bartelmann \& Schneider 1994 and references therein; Ben\'\i tez \&
Mart\'\i nez-Gonz\'alez 1997), we
consider these alternatives unlikely as there appears
no plausible alternative to lensing to explain the
correlations. Nevertheless, if we discard the lensing interpretation,
then the lines-of-sight to the QSOs would be unbiased,
and the shear in PKS1508$-$05 would be the first detection of `cosmic
shear' along an unbiased line-of-sight. 

One has to be careful at this point to clarify the meaning of `cosmic
shear'. In the early work on this subject (e.g., Blandford et al.\ts
1991; Kaiser 1992), the linear evolution of the power spectrum was
considered, and the `cosmic shear' considered was solely due to the diffuse,
large-scale matter distribution. It was predicted to be coherent over scales
below one degree and to have an rms value on arcminute scales of about $1\%$,
depending on the cosmological model. These estimates were revised
upwards by considering the fully non-linear evolution of the power
spectrum (Jain \& Seljak 1996), predicting an rms shear on scales of a
few arcminutes of a few percent. It is unclear at the moment whether
the rms shear on these scales is dominated by fairly large, slightly
non-linear density concentrations, or by fully non-linear collapsed
objects like clusters. In the second case, the shear field around
clusters would have to be included as part of `cosmic shear' as well, 
though a cluster field would constitute a strongly biased
line-of-sight. The detection of a
significant shear in the field of PKS1508$-$05 would then still be
exciting because this would be the first detection of coherent shear in a
direction not targeted towards a known mass concentration.

(2) If the QSOs are indeed magnification biased, the mass responsible
for the observed shear may be related to the mass magnifying the
QSOs. In that case, the detection of cosmic shear in the field of
PKS1508$-$05 would not be along an unbiased direction, but still not
targeted towards a known mass concentration. The detected shear would
then support the magnification bias hypothesis. The true cosmic shear
amplitude would probably be smaller than the value measured by us. 

(3) The shear is caused by material physically associated with the QSO
PKS1508 ($z=1.19$). This would imply that the faint galaxies have a
high-redshift tail which is not implausible, given that the high
redshift cluster MS1054 ($z=0.83$) shows a clear weak lensing signal 
(Luppino \&
Kaiser 1997; see also Deltorn et al. 1997), and also from the number
density depletion in the 
cluster 0024+16 (Fort, Mellier \& Dantel-Fort 1996). In that case the
shear in the field of PKS1508 would be analogous to the one in the
field of 3C324 (Smail \& Dickinson 1996). However, the observations in
the field of 3C324 went to considerably fainter magnitudes and thus
presumably higher mean redshifts of the galaxies. Also, the shear
pattern in PKS1508 seems to be quite uniform, whereas in 3C324 a
systematic tangential alignment relative to the radio source is
seen. We therfore consider material associated with the QSO to be an
unlikely candidate for producing the observed shear. 

In the field of PKS1508, the shear amplitude on a scale of $1'$ is about 
$3\%$, and thus in approximate agreement with theoretical
expectations. However, 
our results do not allow for any quantitative 
cosmological interpretation for two
principal reasons. First, if one considers the possibility (2) as the most
likely one, the shear was measured along a biased line-of-sight and
may therefore not be representative of a random line-of-sight. Second,
in order to draw any conclusion from cosmic shear measurements,
several uncorrelated lines-of-sight have to be observed (e.g., Kaiser
1996). In the near future measurements along unbiased lines-of-sight
with much larger fields of view will be possible. The key 
factor in obtaining quantitative results on weak lensing 
will be the control of systematic distortions in the
instruments. Our results demonstrate that the stability of the 
PSF of the SUSI camera across its field could compensate for the 
relatively small size of the field by providing shear measurements 
of very high significance. 

\medskip
\sec{Acknowledgements}
\SFB
\def\ref#1{\vskip1pt\noindent\hangindent=40pt\hangafter=1 {#1}\par}
\sec{References}
\ref{Bartelmann, M.\ 1995, A\&A 298, 661.}
\ref{Bartelmann, M. \& Schneider, P.\ 1992, A\&A 259, 413.}
\ref{Bartelmann, M. \& Schneider, P.\ 1994, A\&A 284, 1.}
\ref{Ben\'\i tez, N. \& Mart\'\i nez-Gonz\'alez, E.\ 1997,
ApJ 477, 27.}
\ref{Bernardeau, F., van Waerbeke, L. \& Mellier, Y.\ 1996,
A\& A {\it in press}, astro-ph/9609122.} 
\ref{Bertin, E.\ 1996, PhD Thesis. Universit\'e Paris VI.}
\ref{Bertin, E.\& Arnouts, S. 1996, A\&AS 117, 393.}
\ref{Blandford, R.D., Saust, A.B., Brainerd, T.G. \& Villumsen, J.V.\
1991, MNRAS 251, 600.}
\ref{Bonnet, H. \& Mellier, Y.\ 1995, A\&A 303, 331.}
\ref{Bower, R. \& Smail, I. \ 1997. Preprint astro-ph/9612151.} 
\ref{Deltorn, J.-M., Le F\`evre, O., Crampton, D. \& Dickinson, M.\
1997, astro-ph/9704086.}
\ref{Fahlman, G., Kaiser, N., Squires, G. \& Woods, D.\ 1994, ApJ 437, 56.}
\ref{Fort, B. \& Mellier, Y.\ 1994, A\&AR 5, 239.}
\ref{Fort, B., Mellier, Y., Dantel-Fort, M., Bonnet, H. \& Kneib,
J.-P.\ 1996, A\&A 310, 705 (FMDBK).}
\ref{Fort, B., Mellier, Y. \& Dantel-Fort, M.\ 1996, astro-ph/9606039.}
\ref{Jain, B. \& Seljak, U.\ 1996, astro-ph/9611077.}
\ref{Kaiser, N.\ 1992, ApJ 388, 272.}
\ref{Kaiser, N.\ 1996, astro-ph/9610120.}
\ref{Kaiser, N. \& Squires, G.\ 1993, ApJ 404, 441.}
\ref{Kaiser, N., Squires, G. \& Broadhurst, T.\ 1995, ApJ 449, 460.}
\ref{Luppino, G. \& Kaiser, N.\ 1997, ApJ 475, 20.}
\ref{Miralda-Escud\'e, J.\ 1991, ApJ 380, 1.}
\ref{Mould, J.\ et al.\ 1994, MNRAS 271, 31.}
\ref{Narayan, R. \& Bartelmann, M.\ 1996,  astro-ph/9606001.}
\ref{Schneider, P.\ 1996, in: {\it The universe at
high-$z$, large-scale structure and the cosmic microwave
background}, Proceedings of an advanced summer school, Laredo,
Cantabria, Spain,  eds. E. Mart\'\i nez-Gonzales and J.L. Sanz, Lecture
Notes in Physics 470, Springer-Verlag, p.\ 148.}
\ref{Seitz, C., Kneib, J.-P., Schneider, P. \& Seitz, S.\ 1996, A\&A
314, 707.}
\ref{Smail, I. \& Dickinson, M.\ 1995, ApJ 455, L99.} 
\ref{Squires, G.\ et al.\ 1996, ApJ 461, 572.}
\ref{Steidel, C., Dickinson, M., Meyer, D., Adelberger, K. \& Sembach,
K.\ 1996, astro-ph/9610230.}
\ref{Tyson, J.A.\ 1986, AJ 92, 691.}
\ref{van Waerbeke, L.\ 1997 PhD Thesis Universit\'e Paris XI Orsay}
\ref{van Waerbeke, L., Mellier, Y., Schneider, P., Fort, B. \& Mathez,
G.\ 1997, A\&A 317, 303.}
\ref{van Waerbeke, L. \&  Mellier, Y.\ 1997, proceedings of the
XXXIi\`emes Rencotres de Blois. Les Arc. Astro-ph/9606100.}
\ref{Villumsen, J.\ 1996a, MNRAS 281, 369.}
\ref{Villumsen, J.\ 1996b, MNRAS, submitted.}
\ref{White, S.D.M., Navarro, J.F., Evrard, A.E. \& Frenk, C.S.\ 1993,
Nat 366, 429.}
\ref{Wilson, G., Cole, S. \& Frenk, C.S.\ 1996, MNRAS 280, 199.}
\vfill\eject\end